\documentstyle [12pt,aaspp4]{article}
\def\beq{\begin{equation}}
\def\eeq{\end{equation}}
\def\bey{\begin{eqnarray}}
\def\eey{\end{eqnarray}}

\def\cm{\,{\rm {cm}}}
\def\mpc{\,{\rm {Mpc}}}
\def\kpc{\,{\rm {kpc}}}
\def\kpch{\,{h^{-1}{\rm kpc}}}
\def\mpch{\,h^{-1}{\rm {Mpc}}}
\def\kms{\,{\rm {km\, s^{-1}}}}
\def\msun{M_\odot}
\def\vcir{V_c}
\def\v200{V_{200}}
\def\neff{n_{\rm eff}}
\def\rh{r_{200}}
\def\Rd{R_d}
\def\Mh{M}
\def\Md{M_d}
\def\Mb{M_b}
\def\md{m_d}
\def\mb{m_b}
\def\mag{\,\rm mag}

\def\Ld{L_d}
\def\mH{m_{\rm H}}
\def\mtold{\Upsilon _d}
\def\lsun{L_\odot}
\def\Eh{E}
\def\Jh{J}
\def\Jd{J_d}
\def\fd{m_d}

\def\fB{f_B}

\def\jd{j_d}
\def\fv{f_V}
\def\fr{f_R}
\def\tCDM{\rm \tau CDM}
\def\LCDM{\rm \Lambda CDM}
\def\nhi{N_{\rm HI}}
\def\omnow{\Omega_0}
\def\obnow{\Omega_{b,0}}
\def\deltac{\delta_c}
\def\delch{\delta_{0}}

\def\ovnow{\Omega_{\rm \Lambda,0}}
\def\rhoc{\rho_{\rm crit}}
\begin{document}
\title	{The Formation of Galactic Disks}
\author	{H.J. Mo, Shude Mao and Simon D.M. White}
\affil {Max-Planck-Institut f\"ur Astrophysik
\\Karl-Schwarzschild-Strasse 1, 85748 Garching, Germany}
\slugcomment{submitted to MNRAS}
\received{.............}
\accepted{.............}

\begin{abstract}

We study the population of galactic disks expected 
in current hierarchical clustering models for
structure formation. A rotationally supported disk with 
exponential surface density profile is assumed to form
with  a mass and angular
momentum which are fixed fractions of those of its
surrounding dark halo. We assume that haloes respond
adiabatically to disk formation, and that
only stable disks can correspond to real systems. 
With these assumptions the predicted population can 
match both present-day disks and the damped Ly$\alpha$ 
absorbers in QSO spectra. Good agreement is found provided:
(i) the masses of disks are a few percent of those of their haloes;
(ii) the specific angular momenta of disks are similar to
those of their haloes; (iii) present-day disks were 
assembled recently (at $z\leq 1$). In particular, the observed
scatter in the size-rotation velocity plane is reproduced,
as is the slope and scatter of the Tully-Fisher relation.
The zero-point of the TF relation is matched for a stellar
mass-to-light ratio of 1 to 2 $h$ in the I-band, consistent
with observational values derived from disk dynamics.
High redshift disks are predicted to be small and dense, and
could plausibly merge together to form the observed population of
elliptical galaxies. In many (but not all) currently popular
cosmogonies, disks with rotation velocities exceeding 
200 km/s can account for a third or more of the observed 
damped Ly$\alpha$ systems at $z\sim 2.5$.
Half of the lines-of-sight to such systems
are predicted to intersect the absorber at $r\ga 3\kpch$ 
and about 10\% at $r>10\kpch$. The cross-section for absorption is 
strongly weighted towards disks with large angular momentum
and so large size for their mass. The galaxy population
associated with damped absorbers should thus be biased towards
low surface brightness systems.
\end{abstract}

\keywords {galaxies: formation - galaxies: structure - galaxies: spiral
- cosmology: theory - dark matter }

\section {INTRODUCTION}

An important goal of cosmology is to understand how galaxies
form. In standard hierarchical models dissipationless 
dark matter aggregates into larger and larger clumps as 
gravitational instability amplifies the weak density perturbations 
produced at early times. Gas associated with such dark haloes cools 
and condenses within them, eventually forming 
the galaxies we see today. These two aspects of galaxy formation 
are currently understood at very different levels. Gravitational 
processes determine the abundance, the internal structure and 
kinematics, and the formation paths of the dark haloes, can be 
simulated in great detail using N-body methods. Well-tested 
analytic models are now available for describing these 
properties of the halo population. The growth of the dark haloes is 
not much affected by the baryonic components, but determines how 
they are assembled into nonlinear units. 
Gas cools and collects at the centre of each dark halo
until it produces an independent self-gravitating unit which can
form stars, heating and enriching the rest of the gas, and
perhaps even ejecting it from the halo. These
star-formation and feedback processes are not understood in detail 
and can be modeled or simulated only crudely. Furthermore,
as haloes merge to form groups and clusters of galaxies, the cluster
members interact and merge both with each other and with the 
intracluster gas. It will not soon be possible to simulate
galaxy formation {\it ab initio}.
At least the star-formation and feedback processes must be put in 
{\it ad hoc} based on observations of real galaxies.

In a first exploration of the galaxy population expected in this hierarchical 
picture White \& Rees (1978) calculated the expected luminosity
function. The observed characteristic luminosity of galaxies could be 
reproduced provided (i) that feedback prevents efficient conversion of
gas into stars in early objects, and (ii) that merging of galaxies is 
inefficient as groups and clusters of galaxies build up. The models
appeared to overproduce faint galaxies. The formation of galaxy disks
within this scheme was studied by Fall \& Efstathiou (1980). They
showed that if galactic spin is produced by tidal torques acting 
at early times, then extended massive haloes are 
necessary if disks as large as those observed are to form by the 
present day. They also found that plausible initial angular momentum
distributions for the gas could lead to
near-exponential density profiles for the final disk. More recent
analytic work has refined the calculation of global properties
of galaxies, for example the distributions of luminosity, colour,
morphology and metallicity can all be calculated explicitly in
currently popular cosmogonies (e.g. White \& Frenk 1991;
Kauffmann, White \& Guiderdoni 1993; Cole et al 1994; Baugh et al
1997). There has, however,  been relatively little work on the internal
structure of galaxies. Kauffmann (1996b) studied the accumulation, 
star-formation and chemical enrichment histories of individual disks, 
linking them to damped Ly$\alpha$ absorbers at high redshift, but she 
did not consider the scatter in angular momentum for disks 
of a given mass and the resulting scatter in their present-day
properties. Dalcanton, Spergel \& Summers (1997, hereafter DSS) considered
the latter issue in some detail, but did not tie their model 
explicitly to its cosmogonical context or consider its implications 
for disk evolution. There have also been surprisingly few 
well-resolved simulations of galaxy formation from initial conditions which
properly reflect the hierarchical model. So far all have failed
to produce viable models for observed spirals because excessive 
transfer of angular momentum from gas to dark matter results in
overly small disks (e.g. Navarro \& Steinmetz 1997).

In the present paper, we again address the problem of disk formation in
hierarchical cosmogonies, formulating a simple model based on four
key assumptions: 1) the mass of a disk is some fixed fraction of that
of the halo in which it is embedded; 2) the angular momentum of the
disk is also a fixed fraction of that of its halo; 3) the disk
is a thin centrifugally supported structure
with an exponential surface density profile; and 4) only dynamically
stable systems can correspond to real galaxy disks. We derive the
abundance of haloes in any specific cosmogony from the Press-Schechter
formalism, we model their density profiles using the precepts of
Navarro, Frenk \& White (1996, 1997 hereafter NFW), and we assume
their angular momentum distribution to have the near-universal
form found in N-body simulations. (See Lacey \& Cole 1996 for
a summary of how well these analytic models fit simulation results.)
Our four assumptions are then sufficient to predict the properties of
the disk population, once values are assumed for the mass and angular 
momentum fractions and the redshift of disk formation. A mass-to-light
ratio for the disk stars must also be assumed in order to compare with the
luminosities of observed disks. Our assumptions are similar to
those of DSS but
differ in detail. We calculate many of the same properties as they
do, but in addition we address broader issues such as the evolution
of disks, the dependence of the disk population on cosmological
parameters, and the link with QSO absorption lines. 

The outline of our paper is as follows. Section 2 presents our
modelling assumptions in detail. To clarify the scaling of disk properties with
model parameters, we first treat haloes as singular isothermal 
spheres and neglect the gravitational effects of the disks. We then 
consider the more realistic NFW
profiles and include the gravitational effects of the disks. 
In Sections 3 to 5, we describe 
the predicted properties of present-day disks, we discuss TF relations, 
we compare high redshift disks with QSO absorption line systems,
and we study the effect of a nonrotating central bulge.  
Finally in Section 6, we discuss other issues
related to our model. 

\section{MODELS FOR DISKS IN HIERARCHICAL COSMOGONIES}

\subsection{The cosmogonies}

In cold dark matter cosmogonies the evolution of the dark
matter is determined by the parameters of the background cosmological
model and by the power spectrum of the initial density fluctuations.
The relevant cosmological parameters are the Hubble
constant $H_0=100h\kms\mpc ^{-1}$, the total matter density $\omnow$,
the mean density of baryons $\obnow$, and the cosmological constant
$\ovnow$. The last three are all expressed in units of the critical
density for closure. The power spectrum $P(k)$ is specified in CDM
models by an amplitude, conventionally quoted as $\sigma_8$, the
{\it rms} linear overdensity at $z=0$ in spheres of radius $8h^{-1}$
Mpc, and by a characteristic scale $\Gamma=\omnow h/R$, where $R$ is
the current radiation density in units of the standard value
corresponding to a $T=2.73$K black-body photon field and
equilibrium abundances of three massless neutrinos. For
such cosmogonies the abundance of haloes as a function of mass and 
redshift can be calculated quite accurately using the Press-Schechter
formalism, while their radial density profiles can be calculated as a
function of these same variables using the results of Navarro et al
(1997). We give some details of how we implement these formalisms in the
Appendix. 

In the main body of this paper we will compare results for four
different CDM cosmogonies. These are specified by particular values of
the parameter set ($\omnow, \ovnow, h, \Gamma, \sigma_8$). We do
not explicitly need to assume a value for the baryon density $\obnow$,
although for consistency we should require $\obnow> m_d\omnow$ where
$m_d$ is the ratio of disk mass to halo mass adopted (and assumed
universal) in
our models. This ensures that the disk mass is less than the total
initial baryon content of the halo. Our four cosmogonies are:
\begin{itemize}
\begin{enumerate}
\item{the standard cold dark matter model (SCDM), with 
$\omnow=1$, $\ovnow=0$, $h=0.5$, $\Gamma=0.5$, and $\sigma_8=0.6$;}
\item{a COBE-normalized cold dark matter model (CCDM), which
has the same parameters as SCDM, except that $\sigma_8=1.2$;}
\item{a $\tCDM$ model, with $\omnow=1$, $\ovnow=0$, $h=0.5$, 
$\Gamma=0.2$, and $\sigma_8=0.6$; and}
\item{a $\LCDM$ model, with $\omnow=0.3$, $\ovnow=0.7$, $h=0.7$, 
$\Gamma=0.2$, and $\sigma_8=1.0$.}
\end{enumerate}
\end{itemize}
The $\tCDM$ model could be realised if the
$\tau$-neutrino were unstable to decay into other neutrinos and had
a mass and lifetime satisfying $m_\tau t_\tau\approx 17~$keV~yr
(Efstathiou et al 1992). On the relevant scales the power spectrum in
this model is quite similar to that of a mixed dark matter (MDM) model with 
20 per cent of the critical density in stable neutrinos.
The values of $\sigma_8$ adopted for $\tCDM$ and $\LCDM$
are consistent both with the observed abundance
of rich clusters (see e.g. White, Efstathiou \& Frenk 1993;
Mo, Jing \& White 1996) and with observed anisotropies in the cosmic 
microwave background (see e.g. White \& Scott 1996),
whereas the value for SCDM is consistent only
with the cluster normalization and that for CCDM only
with the COBE normalization.  

\subsection {Non-self-gravitating disks in isothermal spheres}

Many of the properties we predict for galaxy disks can
be understood using a very simple model in which dark haloes are
treated as singular isothermal spheres and the gravitational effects
of the disks themselves are neglected. We will examine the limitations
of this rather crude treatment in section 2.3.

For a singular isothermal
sphere the density profile is just,
\beq
\rho (r)={\vcir ^2\over 4\pi G r^2},
\eeq
where the circular velocity $\vcir$ is independent of radius.
Based on the spherical collapse model (Gunn \& Gott 1972; Bertschinger
1985; Cole \& Lacey 1996), we define
the limiting radius of a dark halo to be the radius
$\rh$ within which the mean mass
density is $200\rhoc$, where $\rhoc$ is the critical density
for closure at the redshift $z$ when
the halo is identified. Thus, the radius and mass of a halo
of circular velocity $\vcir$ seen at redshift $z$ are
\beq  \label{rh_h}
\rh ={\vcir \over 10 H(z)};\,\,\,\,\,\,\,
\Mh={\vcir ^2\rh \over G}={\vcir ^3 \over 10 G H(z)} ,
\eeq 
where 
\beq \label{hubblec}
H(z)=H_0\left[\ovnow +(1-\ovnow-\omnow)(1+z)^2+\omnow
(1+z)^3\right]^{1/2}
\eeq
is the Hubble constant at redshift $z$.  
The redshift dependence of halo properties, and so of the disks
which form within them, is determined by $H(z)$. In order to understand
this dependence better, Figure 1 shows $H(z)/H_0$ as a 
function of $z$ for open and flat cosmologies  
with various $\omnow$. $H(z)$ increases
more rapidly with $z$ in universes with larger $\omnow$.   
For a given $\omnow$, the increase is more rapid
for an open universe than for a flat universe. 
At $z\ga 1$, the ratio $H(z)/H_0$ is about twice as large 
in an Einstein-de Sitter universe as in a flat universe 
with $\omnow=0.3$. As a result of this difference,  
the properties of high redshift disks will depend 
significantly on $\omnow$ and $\ovnow$.

We assume that the mass which settles into the disk 
is a fixed fraction $\md$ of the halo mass. The disk mass is then
\beq \label{md_sis}
\Md= {\fd\vcir ^3\over 10 G H(z)}
\approx 1.7\times 10^{11} h^{-1}\msun
\left({\fd\over 0.05}\right)\left({\vcir\over 250\kms}\right)^3
\left[{H(z)\over H_0}\right]^{-1}.
\eeq
Except in Section 5 where we investigate the effects of
including a central bulge, we distinguish only halo and disk components,
and we assume the disks to be thin, to be in centrifugal balance, and
to have exponential surface density profiles,
\beq  \label{ex_disk}
\Sigma (R)=\Sigma_0 \exp\left(-R/\Rd\right).
\eeq
Here $\Rd$ and $\Sigma_0$ are the disk scalelength and
central surface density, and are related to the disk mass through
\beq \label{srd}
\Md = 2\pi \Sigma_0 \Rd ^2.
\eeq
If the gravitational effect of the disk is neglected, its rotation
curve is flat at the level $\vcir$ and its angular momentum is just
\beq \label{jdtwo_sis}
\Jd=2\pi \int \vcir \Sigma (R) R^2  d R=4\pi \Sigma_0 \vcir \Rd^3 =
2M_dR_d\vcir .
\eeq
We assume this angular momentum to be a fraction $\jd$ of
that of the halo, i.e.
\beq \label{jdone_sis}
\Jd=\jd \Jh,
\eeq
and we relate $\Jh$ to the spin parameter $\lambda$ of the halo
through the definition
\beq \label{jh_sis}
\lambda=\Jh\vert \Eh\vert ^{1/2} G^{-1} \Mh^{-5/2},
\eeq
where $\Eh$ is the total energy of the halo.
Equations (\ref{jdtwo_sis})-(\ref{jdone_sis}) then imply that
\beq \label{rdone_sis}
\Rd={\lambda G\Mh ^{3/2}\over 2\vcir \vert \Eh\vert ^{1/2}}
\left({\jd\over \md}\right).
\eeq 
The total energy of a truncated singular isothermal sphere is easily obtained from the
virial theorem by assuming all particles to be on circular
orbits:
\beq \label{eh_sis}
\Eh=-{G\Mh^2\over 2\rh} = -{M\vcir^2\over 2}.
\eeq
On inserting this into equation (\ref{rdone_sis}) 
and using equations (\ref{rh_h}) and (\ref{srd})
we get
\beq \label{rd_sis}
\Rd={1\over \sqrt {2}} 
\left({\jd\over \md}\right)
\lambda \rh
\approx 8.8 h^{-1}\kpc \left({\lambda \over 0.05}\right)
\left({\vcir \over 250\kms}\right)
\left[{H\over H_0}\right]^{-1}\left({\jd\over \md}\right),
\eeq
and
\beq \label{sig_sis}
\Sigma_0\approx 4.8\times 10^{22}h~\cm ^{-2}
\mH \left({\fd\over 0.05}\right)
\left({\lambda\over 0.05}\right)^{-2}
\left({\vcir \over 250\kms}\right)
\left[{H(z)\over H_0}\right]\left({\md\over \jd}\right)^2,
\eeq
where $\mH$ is the mass of a hydrogen atom. 
Using equations (\ref{ex_disk}), (\ref{rd_sis}) and (\ref{sig_sis}), 
we can also 
obtain the limiting radius, $R_l$, at which the disk
surface density drops below some critical hydrogen column $N_l$ (which
must be less than $\Sigma_0/\mH$):
\beq \label{rl}
R_l\approx \Rd \left\lbrace 5.5+ 
\ln \left[\left({\fd\over 0.05}\right)
\left({\md\over \jd}\right)^2
\left({\lambda\over 0.05}\right)^{-2}
\left({\vcir \over 250\kms}\right)
{H(z)\over H_0}
\left({N_l\over 2\times 10^{20}h~\cm ^{-2}}\right)^{-1}
\right]\right\rbrace .
\eeq 
This radius depends on $\rh$, $\lambda$ 
and $\jd/\md$ primarily through the scalelength, $\Rd$.

Equations (\ref{md_sis}) and (\ref{rd_sis})-(\ref{rl}) give the scalings 
of disk properties with respect to a variety of physical parameters. 
These properties are completely 
determined by the values of $\vcir$, $\md$, $\jd$, $\lambda$ and $H(z)$;
other cosmological parameters, such as $z$, $\omnow$ and $\ovnow$, affect disks 
only indirectly through $H(z)$. Since $H(z)$ increases with $z$,
disks of given circular velocity are less massive, smaller and have higher
surface densities at higher redshifts. 
At a given redshift, they are larger and less compact in haloes 
with larger $\lambda$, because they contract less before coming to
centrifugal equilibrium. We will approximate the distribution
of $\lambda$ by
\beq \label{plambda}
p(\lambda) \,d\lambda
={1\over \sqrt{2\pi} \sigma_\lambda}
\exp \left[-{\ln ^2 (\lambda/{\overline \lambda})
\over 2 \sigma_\lambda^2}\right] {d\lambda \over \lambda},
\eeq
where ${\overline \lambda} = 0.05$ and $\sigma_\lambda=0.5$.
This function is a good fit to the N-body results
of Warren et al (1992; see also Cole \& Lacey 1996;
Steinmetz \& Bartelmann 1995) except possibly at very small $\lambda$;
note that it is narrower than the distribution used by DSS.
Since the dependence of $p(\lambda)$ on $\Mh$ and $P(k)$  
is weak, we will not consider it further. This model for
$p(\lambda)$ is plotted as the dotted curve in Figure 10. 
The distribution peaks 
around $\lambda=0.04$, and has a width of about 0.05.
Its 10, 50 and 90 percent points are near $\lambda= 0.025,$ 0.05,
and 0.1, respectively. Thus we see from equations (\ref{rd_sis}) and
(\ref{sig_sis}) that about 10\% of disks of given circular velocity 
will be more than twice the size of the typical disk, and so have 
less than a quarter the typical surface density, while about 10\% 
will be less than half the typical size, and so have more than four 
times the typical surface density. (The latter 10\% may be unstable 
-- see Section 3.2.)

Disk masses are, of course, directly proportional to $\fd$, the
fraction of the halo mass we assume them to contain. For simplicity we
will assume this fraction to be the same for all disks, but even
in this case it is unclear what value of $\fd$ is appropriate. A
plausible upper limit is the baryon fraction of the Universe as a
whole, $\fB\equiv \obnow/ \omnow$, but the efficiency of 
forming disks could be quite low, implying that $\fd$ could be
substantially smaller than $\fB$. Taking a Big Bang nucleosynthesis 
value for the baryon density, $\obnow=0.015h^{-2}$
(e.g. Walker et al 1991, but see Turner 1996) gives an upper
limit on $\fd$ in the range 0.05 to 0.1 for the 
cosmological models we consider (see Section 2.1).

If the specific angular momentum of the material which forms the disk
is the same as that of the halo, then $\jd/\md=1$. This has been the
standard assumption in modelling disk formation since the work of Fall
\& Efstathiou (1980), but it is unclear if it is appropriate,
particularly if the the efficiency of disk formation, $\fd/\fB$, is
significantly less than unity. Even when this efficiency
is high, numerical simulations of spiral galaxy formation have tended to find 
$\jd/\md$ values well below unity (e.g. Navarro \& White
1994; Navarro \& Steinmetz 1997). Equation
(\ref{rd_sis}) shows that disk sizes are directly proportional to $\jd/\md$,
and we will find below that values close to unity are required to fit
observed spirals. This agrees with the fact that the simulations did indeed
produce galaxies with too little angular momentum.
Most of our discussion will assume $\jd=\md$, but we will comment on
the effects of changing $\md$ and $\jd$ independently.
 
The disk of our own Galaxy has a 
mass of about $6\times 10^{10}\msun$.
Its scalelength and circular velocity (measured at the solar radius, 
$R_\odot\sim 8\kpc$) are $\Rd\sim 3.5\kpc$ and 
$\vcir\sim 220\kms$ (see Binney \& Tremaine 1987, p17). 
Since the Milky Way appears to be a typical Sbc galaxy,
it is interesting to see whether a disk with these parameters is 
easily accommodated. Equation (\ref{md_sis}) together with
Figure 1 implies that Milky Way-like disks 
cannot form at $z> 1$
in a universe with $\omnow\sim 1$; for our nucleosynthesis value of
$\fB$ the disk mass for 
$\vcir=220\kms$ is too small even if disks are assumed to 
form with maximal efficiency. The constraints are weaker 
for a low-$\omnow$ universe, because $\fB$ is larger and
$H(z)$ increases less rapidly with $z$. A similar conclusion can be derived
from $\Rd$. At high redshift the disk scalelength for $\vcir=220\kms$ 
is too small to match the observed value unless $\lambda\gg 0.05$. 
Such high values of $\lambda$ are associated with only a
small fraction of haloes, and so late formation 
is required for the bulk of the disk population.
As we will see later these constraints become stronger in more
realistic models.

If we assume that all disks have the same stellar
mass-to-light ratio,
$\mtold\equiv \Md/L_d$ (in solar units), we can cast equation (\ref{md_sis}) 
into a form similar to the TF relation:
\beq \label{tf_sis}
\Ld=A\left({\vcir\over 250\kms}\right)^\alpha,
\eeq
where $\alpha=3$ is the slope, and
\beq \label{zerop_sis}
A=1.7\times 10^{11} h^{-1}\lsun \mtold ^{-1}      
\left({\fd\over 0.05}\right)
\left[{H(z)\over H_0}\right]^{-1}
\eeq
is the `zero-point'. Notice that $\lambda$ does not appear in
these equations. As a result, all the disks assembled 
at a given time are predicted to lie on a scatter-free TF relation 
despite the fact that there is substantial
scatter in their scale radii and surface densities.
The zero-point of this TF relation can, however, depend 
on the redshift of assembly, being
lower at higher redshifts if the product $\Upsilon(z) H(z)$
is larger. For I-band luminosities the observed value of $\alpha$  
is very close to 3 (Willick et al 1995; 1996;
Giovanelli et al 1997). Since stellar population synthesis models
suggest that $\mtold$ should vary rather little in this band, we can
take this as confirmation of the approximate validity of our assumption
that $\md$ is independent of $\vcir$. The zero-point of the observed
TF relation corresponds to
a value of $A$ of about $5\times 10^{10}h^{-2}\lsun$
(Giovanelli et al 1997). Since the expected value of the disk
mass-to-light ratio is $\mtold\approx 1.7h$ (Bottema 1997, see section 3.4),
comparison with equation (\ref{zerop_sis}) again suggests 
that late assembly epochs are required; for the appropriate values
of $\fd$ the predicted zero-point would be too low
if disks formed at high redshift.

The limiting radius $R_l$ which we defined in equation (\ref{rl}) is chosen
to be relevant to the damped Ly$\alpha$ absorption lines seen in QSO
spectra. Such systems have HI column densities of 
$\nhi\ga 2\times 10^{20}\cm^{-2}$, comparable to those of
present-day galactic disks (see Wolfe 1995 for a review). 
For given $\vcir$ and $\lambda$ disks are smaller at higher
redshifts. As a result the cross-section for damped 
Ly$\alpha$ absorption decreases with $z$. As one can see from
equation (\ref{rl}), the limiting radius for damped absorption 
is roughly $R_l=5.5\Rd$. For 
$\vcir \sim 200\kms$ and $\lambda\sim 0.05$, 
we find $R_l$ between $5$ and $10\kpch$ at $z\sim 3$ for 
$\omnow$ between 1 and 0.3.
Given the distribution of dark haloes as a function of
$\lambda$ and $\vcir$, it is clearly possible to calculate the
total cross-section for damped absorption along a random sightline,
as well as the
distributions of $\lambda$, $\vcir$, $\nhi$ and impact parameter
(i.e. the distance between the
centre of the absorbing disk and the sightline to the
QSO) for the systems actually seen as damped absorbers. We discuss
this in detail in section 4.

The model presented in this subsection is limited both because
the dark haloes formed by dissipationless hierarchical clustering are 
only very roughly approximated by singular isothermal spheres, and because
the gravitational effects of the disks are often not negligible.
Despite this, most of the characteristic properties of this model are
retained with only rather minor modifications in the more realistic 
model which we discuss next.

\subsection {Self-gravitating disks in haloes with realistic profiles}

In a recent series of papers Navarro, Frenk \& White 
(1995; 1996; 1997, hereafter NFW) used high-resolution
numerical simulations to show
that the equilibrium density profiles of dark matter haloes of
all masses in all dissipationless hierarchical clustering cosmogonies can be 
well fit by the simple formula:
\begin{equation} \label{nfw}
\rho(r) = \rhoc {\delch \over (r/r_s)(1+r/r_s)^2},
\end{equation}
where $r_s$ is a scale radius, and $\delch$ is a 
characteristic overdensity. The logarithmic slope of this profile
changes gradually from $-1$ near the centre to $-3$ at large radii.
As before (and following NFW) we define the limiting radius of a
virialised halo, $\rh$, to be the radius within which 
the mean mass density is $200\rhoc$.
The mass within some smaller radius $r$ is then
\beq \label{mh_nfw}
\Mh(r) = 4 \pi \rhoc \delch r_s^3 
\left[{1 \over 1+c x} - 1 + \ln(1+ c x)\right],
\eeq
where $x \equiv r/\rh$, and
\beq \label{concentration}
c \equiv  {\rh \over r_s}
\eeq
is the halo concentration factor. The total
mass of the halo is given by equation (\ref{mh_nfw})
with $x=1$. Using equations (\ref{nfw})-(\ref{concentration}), 
we can obtain a
relation between the characteristic overdensity
and the halo concentration factor,
\beq \label{delta_null}
\delch={200 \over 3} {c^3 \over \ln(1+c)-c/(1+c)}.
\eeq
We can obtain the total energy of the truncated halo as before
by assuming all particles to be on circular orbits,
calculating their total kinetic energy, and then using the virial
theorem. This yields
\beq \label{eh_nfw}
E
= - G { (4 \pi \rhoc \delch r_s^3)^2\over 2r_s}
\left[ {1 \over 2} - {1 \over 2 (1+c)^2} -
{\ln(1+c) \over 1+c} \right] = - {G \Mh^2 \over 2 \rh} f_c,
\eeq
where 
\beq
f_c={c \over 2}~{1-1/(1+c)^2-2\ln(1+c)/(1+c) \over [c/(1+c)-\ln(1+c)]^2}
\approx {2 \over 3} + \left({c \over 21.5}\right)^{0.7}.
\eeq
The approximation for $f_c$ given above is accurate to within 1\% for
the relevant range $5<c<30$ (to about 3\% for $0<c<50$).
Comparing equation (\ref{eh_nfw}) with equation (\ref{eh_sis}), we see that
the total energy for an NFW profile differs from that
for an isothermal sphere by the factor, $f_c$, which depends
only on the concentration factor. 

If the gravitational effects of the disk were negligible, its rotation
curve would simply follow the circular velocity curve of the
unperturbed halo, $\vcir^2(r)=GM(r)/r$. This curve rises to a maximum
at $r\approx 2r_s$ and then falls gently at larger radii (see NFW). In
fact, disk formation alters the rotation curve
not only through the direct gravitational effects of the disk, but
also through the contraction it induces in the inner
regions of the dark halo. We analyse this effect by assuming that the
halo responds adiabatically to the slow assembly of the disk, and
that it remains spherical as it contracts; the angular
momentum of individual dark matter particles is then conserved
and a particle which is initially at mean
radius, $r_i$, ends up at mean radius, $r$, where
\beq \label{action}
G \Mh_{f}(r) r = G \Mh(r_i) r_i.
\eeq
In this formula  $\Mh(r_i)$ is given by the NFW profile (\ref{mh_nfw})
and $\Mh_f(r)$ is the total final mass within $r$. [See Barnes \&
White (1984) for a test of this adiabatic model.]
The final mass is the sum of the dark matter mass
inside the initial radius $r_i$ and the mass contributed by
the exponential disk. Hence
\beq \label{adiabatic}
\Mh_f(r) = \Md(r) + \Mh(r_i) (1-\md),
\eeq
where, as before, $\md$, is the fraction of the total has in the
disk, and
\beq \label{md_r}
\Md(r) = \Md \left[ 1-\left(1+{r\over \Rd}\right) e^{-r/\Rd} \right].
\eeq
Here we implicitly assume that the baryons initially had the same
density profile as the dark matter, and those which do
not end up in the disk remain distributed in the same way as the
dark matter. For a given rotation curve, $\vcir (R)$,
the total angular momentum of the disk can be written as
\beq 
J_d = \int_0^{\rh} \vcir(R) R \Sigma(R) 2 \pi R dR
= \Md \Rd \v200 \int_0^{\rh/R_d}
e^{-u} u^2 {\vcir(\Rd u) \over \v200~} du,
\eeq
where $\v200\equiv\vcir(\rh)$ is unaffected by
disk formation.
In practice we can set the upper limit of integration to infinity
because the disk surface density drops exponentially and
$\rh \gg R_d$. Substituting equation (\ref{eh_nfw}) into 
equation (\ref{jh_sis}) and writing $\Jd=\jd\Jh$, we can use
the argument of Section 2.2 to obtain 
\beq \label{rd_nfw}
\Rd = {1\over \sqrt{2}}\left({\jd\over \md}\right)
\lambda \rh f_c^{-1/2} \fr (\lambda,c,\md,\jd)
\eeq
where 
\beq \label{fr_nfw}
\fr (\lambda,c,\md, \jd) 
= 2 \left[\int_0^\infty
e^{-u} u^2 ~{\vcir(\Rd u)\over\v200}~ du\right]^{-1}.
\eeq
Comparing equation (\ref{rd_nfw}) with equation (\ref{rd_sis}) we 
see two effects that cause the
disk scalelength to differ
from that in a singular isothermal sphere;
the factor $f_c^{-1/2}$ comes from the change in total energy 
resulting from the different density profile, while the factor
$\fr  (\lambda,c,\md, \jd)$ is due both to the different density
profile and to the gravitational effects of the disk. 

The rotation velocity $\vcir(r)$ is a sum in quadrature of
contributions from the disk and from the dark matter:
\beq \label{vc_nfw}
\vcir^2(r) = V_{c,d}^2(r) + V_{c, DM}^2(r),
\eeq
where
\beq \label{vc_bdm}
V_{c, DM}^2(r)=G\left[\Mh_f(r)-\Md (r)\right]/r,
\eeq
and $V_{c,d}(r)$ is the rotation curve which would be produced by the
exponential disk alone. Note that when calculating the latter quantity
the flattened geometry of the
disk has to be taken into account (Binney \& Tremaine 1987, p77).
For a given set of parameters, $\v200$, $c$, $\lambda$, $\md$ and $\jd$,
equations (\ref{adiabatic}), (\ref{md_r}), (\ref{rd_nfw}) 
and (\ref{vc_nfw}) must be solved by iteration to yield the
scale length, $\Rd$, and the rotation curve, $\vcir (R)$.
For example, we can start with a guess for $\Rd$ by setting $f_R=1$
in equation (\ref{rd_nfw}). We can then
obtain $\Md (r)$ from equation (\ref{md_r}). Substituting this
into equation (\ref{adiabatic}) and using equation (\ref{action}),
we can solve for $r_i$ as a function of $r$ and so obtain 
$\Mh_f(r)$ from equation (\ref{adiabatic}). With this 
and with $V_{c,d}^2(r)$ calculated for the assumed 
$\Rd$, we can get the disk rotation curve
from equations (\ref{vc_nfw}) and (\ref{vc_bdm}).
Inserting this into equation (\ref{fr_nfw})
and using equation (\ref{rd_nfw}) we then obtain a new value
for $\Rd$. In practice this iteration converges rapidly, 
so that accurate values for both $\Rd$ and $\vcir (r)$ are easily 
obtained. 

It is useful to have a fitting formula for $\fr$ which allows this
iterative procedure to be avoided. We find that the following
expression has sufficient accuracy:
\beq \label{fr_appr}
\fr \approx
\left({\lambda^\prime\over 0.1}\right)^{-0.06+2.71\fd+0.0047/\lambda^\prime}
(1-3\fd+5.2 \fd^2) (1-0.019 c+0.00025 c^2+0.52/c),
\eeq
where $\lambda^\prime\equiv (\jd/\md)\lambda$. We will see in Section 3.1
that the rotation curves predicted by our procedure typically reach a
maximum near $3\Rd$ and in much of the discussion which follows we
will use the value at this point to characterise their amplitude. As a
result, it is also useful to have a fitting formula for the
dimensionless coefficient $\fv (\lambda ,c,\fd ,\jd )$ in
\beq \label{vcc_nfw}
\vcir(3\Rd ) = \left({G \Mh \over \rh}\right)^{1/2} \fv = \v200 \fv .
\eeq
We find
\beq \label{fv_appr}
\fv \approx 
\left({\lambda^\prime\over 0.1}\right)^{-2.67\fd-0.0038/\lambda^\prime
+0.2\lambda^\prime}
(1+4.35\fd-3.76 \fd^2) {1+0.057 c-0.00034 c^2-1.54/c 
\over [-c/(1+c)+\ln(1+c)]^{1/2}}.
\eeq
The approximations in (\ref{fr_appr}) and (\ref{fv_appr}) are 
accurate to within 15\% for $5<c<30$, $0.02<\lambda^\prime<0.2$ and
$0.02<m_d<0.2$.

\section{THE SYSTEMATIC PROPERTIES OF DISKS}

\subsection{Rotation curves}

According to the model set out in Section 2.3 the shape of a disk's
rotation curve depends on the concentration of its halo, $c$, on the
fraction of the halo mass which it contains, $\fd$, and on its angular
momentum as specified by the parameter combination $\lambda^\prime=
(\jd/\fd)\lambda$. For given values of these three parameters the
amplitude of the rotation curve and its radial scale are set by
any single scale parameter, either for the halo ($\Mh$, $\rh$, or $\v200$)
or for the disk ($M_d$, $R_d$, or $\Sigma_0$). Figure 2 shows examples chosen 
to illustrate how changes in $\fd$, $\lambda^\prime$ and $c$ affect the shape
of the rotation curve of a disk of fixed mass. Curves for other disk
masses can be obtained simply by scaling both axes by $M_d^{1/3}$.
In all cases the rotation curves are flat at radii larger than a few 
disk scalelengths. At smaller radii the shape of the curves
depends strongly on the spin parameter. For small values of 
$\lambda^\prime$ the disk is more compact, and its self-gravity
is more important. The rotation velocity then increases rapidly 
near the centre to a peak near $R\sim 3\Rd$, thereafter 
dropping gradually towards a plateau at larger radii.
For large $\lambda^\prime$, the disk is much more extended 
and its self-gravity is negligible. The rotation velocity then 
increases slowly with radius in the inner regions.

Since the surface density of a disk is roughly
proportional to $(\lambda^\prime)^{-2}$, these results imply a
correlation between disk surface brightness
and rotation curve shape which is very reminiscent of the
observational trends pointed out by Casertano \& van Gorkom (1991).
Our predicted rotation curves also become more peaked at small 
radii for larger values of $\fd$, again because of the larger 
gravitational effect of the disk. Rotation curves as peaked as that 
shown in the lower left panel of Figure 2 are not observed. As we
will discuss below, this is probably because such disks are violently
unstable. Halo concentration affects rotation curves in the
obvious way; more strongly peaked curves are found in more
concentrated haloes. Combining these trends we conclude that
slowly increasing rotation curves are expected for low-mass 
disks in haloes with large $\lambda^\prime$ and small $c$.
Since $c$ is smaller for more massive haloes (see NFW),
we would predict slowly rising rotation curves
to occur preferentially in giant galaxies. In fact, the
opposite is observed. This is almost certainly a result of the
greater influence of the bulge in giant systems. A final trend
which is clear from Figure 2 is that at fixed disk mass more
strongly peaked curves have larger maximum rotation velocities.
A similar trend can be seen in the observational data (e.g.
Persic and Salucci 1991).

\subsection {Disk instability}

The modelling of the last two sections can lead to disks with a
very wide range of properties. Not all of these disks, however, are
guaranteed to be physically realisable. In particular, those in
which the self-gravity of the disk is dominant are likely to be
dynamically unstable to the formation of a bar. There is an extensive
literature on such bar instabilities [see Christodoulou, Shlosman \&
Tohline (1995) and references therein]. For our purposes the most
relevant study is that of Efstathiou, Lake \& Negroponte (1982) 
who used N-body techniques to investigate global 
instabilities of exponential disks embedded in 
a variety of haloes. They found the onset of the bar instability
for stellar disks to be characterised by the criterion,
\beq\label{epsilon_m}
\epsilon_{\rm m}\equiv {V_{\rm max}\over (G\Md/\Rd)^{1/2}}\la 1.1,
\eeq
where $V_{\rm max}$ is the maximum rotation velocity of the disk.   
The appropriate instability threshold for gas disks is lower, 
$\epsilon_{\rm m}\la 0.9$, as discussed in 
Christodoulou et al (1995). 
In our model, 
\beq
\epsilon_{\rm m}\approx {1\over 2^{1/4}}
\left({\lambda^\prime\over \md}\right)^{1/2} 
f_c^{-1/4}f_R^{1/2} f_V.
\eeq
Thus, disks are stable if 
\beq\label{lambda_crit}
\lambda^\prime \ga \lambda^\prime_{\rm crit}=\sqrt{2} 
\epsilon_{\rm m,crit}^2 \md f_c^{1/2}f_R^{-1} f_V^{-2}, 
\eeq
where $\epsilon_{\rm m, crit}\approx 1$ is the critical value 
of $\epsilon _{\rm m}$ for disk stability.
If the effect of disk self-gravity on $\Rd$
and $\vcir$ is weak, then $\lambda^\prime_{\rm crit}\sim\md$.
Figure 3 shows $\lambda^\prime_{\rm crit}$ as a function of $\md$ for
our NFW models. Results are shown for 
$\epsilon_{\rm m,crit}=0.8$, 1, and 1.2, in order to bracket the critical 
values of $\epsilon_{\rm m}$ discussed above. In practice we will normally
adopt $\epsilon_{\rm m, crit}=1$ as a fiducial value.
The dependence of $\lambda^\prime_{\rm crit}$ on halo concentration 
is weak, because
for a given $\md$, the dependences of $V_{\rm max}$ and $\Rd$ on $c$
are opposite.
Since halo mass and assembly time affect $\lambda^\prime_{\rm crit}$
only through $c$, we expect disk stability to be almost independent
of these parameters. As one can see from Figure 3, it is a useful rough
approximation to take $\lambda^\prime > \md$ as the condition for
stability. Thus in Figure 2 the top two disks are marginally stable, the
lower left disk is strongly unstable and the lower right disk is stable.
 
Disk galaxies are common and appear to be stable. If we take account 
of the expected distribution of $\lambda$ (equation \ref{plambda})
and the likelihood that $\jd\leq\md$, then the results in
Figure 3 suggest that $\md$ values of 0.05 or
less will be needed to ensure that most haloes host stable
disks rather than the descendants of unstable disks.

\subsection {Disk scalelengths and formation times}

For a given cosmogony our assumptions determine the joint distribution
of disk size and rotation speed once
values are adopted for the mass and angular momentum fractions, $\fd$ and
$\jd$, and for the disk formation redshift. This distribution can be 
compared directly with observation
provided our size measure, $\Rd$, can be identified with
the exponential scalelength of observed luminosity profiles.
This seems reasonable since there is no evidence that the
mass-to-light ratio of the stellar populations in real disks are a
strong function of radius. In our simplified model the disk ``formation
redshift'' is just the time at which the arguments of Section 2
are applied. We make no attempt to follow the actual formation and
evolution of disks, and we assume that all disks form at the same
time. This formation redshift should be interpreted as the epoch when
the material actually observed was assembled into a single virialised
object.

Figure 4 compares predicted distributions of $\Rd$ as a function 
of $\vcir$ with observational data for a large sample of nearby
spirals. The four panels give predictions for formation redshifts of 0
and 1, and for two different cosmogonies, SCDM and $\LCDM$;
predictions for CCDM and $\tCDM$ are similar to those for SCDM.
As a characteristic measure of rotation speed for our model
galaxies we use the value of $\vcir$ at $3\Rd$. As discussed in \S
2.3 this is always close to the maximum of the rotation curve. 
The solid line in each panel is the $\Rd$-$\vcir$ relation
for {\it critical} disks (i.e. those with $\lambda^\prime=
\lambda^\prime_{\rm crit}$) for the case $\md=0.05$; stable disks must
lie above this line. Since $\lambda^\prime_{\rm crit} \approx 0.05$ 
in this case, and since, in general, we expect $\jd\leq\fd$,
implying $\lambda^\prime\leq\lambda$, at most 
half of all dark haloes will contain stable disks 
for $\fd\geq 0.05$ (see the discussion around
equation \ref{plambda}). The short-dashed line in each panel 
is the corresponding critical line for $\md=0.025$. In this case
about 90 percent of  the dark haloes can host stable disks under
the optimistic assumption that $\jd=\fd$. The critical lines would
be even lower if $\md<0.025$, but values of $\lambda$
below 0.025 are rare, and so such compact disks will not be abundant
unless $\jd$ is often substantially smaller than $\fd$.
Finally, the long-dashed lines in Figure 4 show $\Rd$-$\vcir$
relations for $\md=\jd=0.05$ and $\lambda=0.1$. For these parameters
disk self-gravity is negligible and the line is almost unaltered for
$\fd=\jd=0.025$. Fewer than 10 percent of the dark haloes
have $\lambda\ge 0.1$ in the $\lambda$-distribution of
equation (\ref{plambda}). Thus for $\jd\leq\fd$ very few disks are
expected to lie above these long-dashed lines. The slopes and relative
amplitudes of all the theoretical lines in Figure 4 can easily be
understood from the scalings in equation (\ref{rd_sis}).

The data points in Figure 4 are the observational results of Courteau
and coworkers (Courteau 1996; 1997; Broeils \& Courteau 1996) for  
{\it present-day} normal spirals. For a formation redshift
of zero, the observed distribution lies comfortably below the
long dashed lines and above the short dashed lines in both
cosmologies. A significant fraction of the disks lie below the solid
lines, particularly for $\LCDM$, suggesting that these disks must have 
$\fd<0.05$ in order to be stable. For this formation redshift
it seems relatively easy to reproduce the observed
distribution in both cosmogonies. For a formation redshift of 1,
in contrast, the observed points clearly lie too high
relative to the predictions of the SCDM model (and similarly for
CCDM and $\tCDM$). Observed disks are too big to have
formed at $z\sim 1$ in a high density universe, unless there is
some way for $\jd$ to be substantially larger than $\fd$. The
observed points also lie rather high for $z=1$ in $\LCDM$, but here
the discrepancy is marginal provided that $\jd\sim\md$ is indeed 
a realistic expectation. It is clear that substantial transfer of
angular momentum
from baryons to dark matter, resulting in $\jd\ll\fd$, would lead to
unacceptably small disks for any assembly redshift in all of the
cosmogonies we consider. Note that although the observed disks
of Figure 4 apparently formed late, this does not mean that disks were
not present in large numbers at high redshift; early disks
should, however, be significantly smaller than present-day objects of
the same $\vcir$. 

  For the cosmogonies considered here, the comoving number density 
of galaxy-sized haloes is expected to peak at $z\sim 1$
and then to decline gradually at lower redshift.
The peak density is comparable to the number density
of galaxies observed today. Thus, as can be seen explicitly from more detailed
semi-analytic modelling (e.g. Kauffmann et al 1993) these models do
indeed predict enough haloes to house the observed population of
disks. Late formation thus appears viable both in terms of the
structural properties and the abundance of disks in CDM-like models.    

\subsection {Disk surface densities}

For a given cosmogony the distributions of $\Mh$ 
(the halo mass) and $\lambda$ are known as a function of redshift 
[see equations (\ref{ps_mf}) and (\ref{plambda})], and furthermore 
the halo concentration factor, $c$, can be calculated from
$\Mh$ and $z$ (we neglect any scatter in $c$, see the Appendix).
For any particular
formation redshift, we can therefore generate Monte-Carlo samples 
of the halo distribution in the $\Mh$-$\lambda$ plane, and, using specific
values for $\fd$ and $\jd$, transform these into Monte-Carlo samplings
of the disk population. Figure 5 shows the distribution of disks in the  
$\vcir$-$\Sigma_0$ plane for SCDM with $\md=0.025$, 0.5 and 0.1,
$\jd=\md$ and a formation redshift of zero. 
Unstable disks with $\epsilon_{\rm m}<1$ are excluded. 
The instability criterion leads to an upper cutoff in the central
surface luminosity density $\Sigma_0$. For the simple isothermal sphere model
of \S 2.2  this upper cutoff can be written as
\beq
\Sigma_{\rm crit}\propto {\md\over (\lambda^\prime_{\rm crit})^2}
H(z)\vcir\propto {\vcir\over \lambda^\prime_{\rm crit}}H(z)
\propto {\vcir\over \md}H(z),
\eeq
where we have used $\lambda^\prime_{\rm crit}\approx \md$ (see
equation \ref{lambda_crit}). Thus $\Sigma_{\rm crit}$
is larger for higher $z$, for higher $\vcir$ and for lower $\md$.
The last two dependences are clearly seen
in Figure 5.

To compare the distributions of Figure 5 with real data we need to
assume a stellar mass-to-light ratio for the disks. Throughout this
article we will use the values found for high surface brightness disks
by Bottema (1997). From disk dynamics, he inferred $\Upsilon_B = 
(1.79\pm 0.48)$ for $h=0.75$. Since $B-I=1.7$ for a typical disk galaxy
(McGaugh \& de Blok 1996), and
$(B-I)_\odot=1.33$, we obtain $\Upsilon_I=(1.7 \pm 0.5) h$.
Solid squares in Figure 5
correspond to median values for the Courteau data in Figure 4. 
We have converted from
$\vcir$ to $L_I$ using the Tully-Fisher relation of Giovanelli et al
(1997; equation \ref{zero-point}), and from $L_I$ to $\Sigma_0$ using
equation (\ref{srd}) and $\Upsilon_I = 1.7h$. Error bars join the
upper and lower 10\% points of the distribution in each $\vcir$ bin.
The typical central surface brightnesses found in this way are
similar to the ``standard'' value advocated by Freeman (1970). For
$\md=0.05$, the median central surface density for disks at $z=0$ in
SCDM appears to be about a factor of 2
too faint compared with the observed value. However, this
discrepancy should not be over-interpreted since
low surface brightness galaxies are probably
underrepresented in Courteau's sample. In addition, the
model prediction has substantial uncertainties. For a given
$V_c$, the predicted disk surface density is proportional to $H(z)$,
and so a formation redshift of 0.5 would increase the predictions by a
factor of 1.8. Furthermore, the upper cutoff of the
disk surface density distribution depends sensitively on the value of
$\epsilon_{\rm m, crit}$ (cf. equation \ref{epsilon_m}). For
a given $\md$, $\Sigma_{\rm cr} \propto\epsilon^{-4}_{\rm m, crit}$; a
20\% decrease in $\epsilon_{\rm m, crit}$ would thus increase the
upper cutoff on $\Sigma_0$ by a factor of two.

Despite these uncertainties, Figure 5 suggests that $\fd\la 0.05$ 
is preferred. Notice that this conclusion is independent
of the $\Rd$-$\vcir$ analysis of the last section which suggested
similar values for $\md$. It is interesting that the abundance of both
high and low surface density galaxies is predicted to increase as
$\fd$ decreases. This is a consequence of the wider range of $\lambda$
values which produce stable disks when $\md$ is small.
There is currently considerable debate about the relative abundance of
low-surface-brightness (LSB) galaxies. Our results confirm those
of Dalcanton et al (1997); hierarchical models can relatively easily
produce a spread which appears broad enough to be consistent with
the observations. We note that an additional uncertainty in any
detailed comparison comes from the fact that the star formation
efficiency in disks may well vary strongly as a function of surface density.
This would further complicate the conversion between disk mass and
disk light.

\subsection {Tully-Fisher relations}

In one of the most complete recent studies of the Tully-Fisher
relation for nearby galaxies Giovanelli et al (1997) put together a
homogeneous set of HI velocity profiles and I-band photometry for a
set of 555 spiral galaxies in 24 clusters. By careful adjustment of
the relative zero-points of the different clusters they derived
a TF relation in the (Cousin) I-band,
\begin{equation} \label{zero-point}
{\cal M}_{I} - 5\log h = -(21.00 \pm 0.02) - (7.68\pm 0.13) (\log W - 2.5),
\end{equation}
where $W$ is the inclination-corrected width of the HI line profile,
and, to a good approximation, is twice the maximum rotation velocity.
When comparing with our models we will assume $W = 2\vcir(3\Rd)$ ( see
\S 2.3). Giovanelli et al find a mean scatter around this 
relation of 0.35 magnitudes, but point out that the scatter is 
actually magnitude-dependent; it increases significantly with 
decreasing velocity width.
A nearly identical Tully-Fisher relation was derived independently
by Shanks (1997), but the zero-point of the TF relation 
in Willick et al (1996) is fainter by about 0.5 mag.

In order to derive luminosities for the disks in our models we need to
specify their stellar mass-to-light ratio. Unfortunately, the
appropriate value for this quantity is quite uncertain; here we will
again use the value $\Upsilon_I=1.7 h$ which Bottema (1997) derived
for the disks of high surface brightness spirals. 
For a given mass-to-light
ratio, the magnitude of a disk galaxy can be calculated as
\begin{equation}\label{magd}
{\cal M}_{I} = {\cal M}_{\odot, I} - 2.5 \log { L_I \over L_{\odot, I}}
=4.15 - 2.5 \log {\Md \over M_\odot} + 2.5 \log (\Upsilon _I),
\end{equation}
where we have used ${\cal M}_{\odot,V}=4.83$, and
$(V-I)_\odot=0.68$ (Bessel 1979, Table II). Note that in the observed
TF relation ${\cal M}_{I}$ is the total absolute magnitude of the
galaxy, whereas for our models we are estimating the absolute
magnitude of the disk alone; we are therefore implicitly assuming that
the bulge luminosity can be neglected for the galaxies under
consideration (but see \S 5 below).

With these assumptions, we can generate a Monte-Carlo disk catalogue,
as in \S 3.4, and plot ${\cal M}_I$ versus $\vcir(3\Rd)$ ($\approx W/2$).
Figure 6 shows such plots for stable disks with a formation redshift
of $z=0$ in SCDM and $\LCDM$. Here we assume $\md=\jd=0.05$.
For each cosmogony, the three panels show results for
three choices of $\epsilon_{\rm m,crit}$, the critical value
of $\epsilon_{\rm m}$ for disk instability (equations \ref{epsilon_m} to 
\ref{lambda_crit}). As one can see, the plots are similar in all cases.
The zero point for the SCDM model is lower than that for the
$\LCDM$ model for the reasons to be discussed in more detail in section 3.5.3.
The scatter is slightly larger for smaller values of
$\epsilon_{\rm m, crit}$, because disks are then stable
for smaller $\lambda$ values where disk self-gravity is more important.
By fitting plots such as these we can clearly derive TF relations for
our models, and so can compare their slope, scatter and zero-point
with those of the observed Tully-Fisher relation.

\subsubsection{Tully-Fisher slopes}

We begin by considering the slopes of the model relations. The solid 
lines in Figure 6 are the linear regressions of 
${\cal M}_I$ on $\vcir$ for each Monte-Carlo sample of simulated data.
It is clear that in all cases the slope is close to 
$\alpha=3$ (where $\alpha$ is defined in equation \ref{tf_sis}), 
and is consistent with the observed value of Giovanelli et al which we
indicate with a dashed line. Although there is clearly very good
agreement, it
is important to realise that this is a consequence of choosing to compare
with the observed I-band data. In fact, the observed slope
is quite dependent on the photometric
band used. For example, in the B-band it
is about $\alpha=2.5$ (Strauss \& Willick 1995 and references therein).
This difference arises because the 
colours of disk galaxies vary systematically with their luminosity;
fainter galaxies show proportionately more star formation and are
bluer. Stellar population models suggest that the stellar
mass-to-light ratios of disks must vary quite strongly with luminosity
in the B-band (being smaller for the fainter, bluer galaxies) but 
could plausibly be constant at I. For this reason the I-band TF
relation does indeed appear to be the most appropriate one to compare 
with our models. 

In the models the slight deviation of the slope from
$\alpha=3$ is caused by the fact that massive haloes are less concentrated
than low-mass haloes. To see this, we recall NFW's discovery that
the concentration factor of a halo is determined by its formation
time $z_f$; massive haloes form later and so are less
concentrated (see the Appendix). Since the value of $f$ in equation 
(\ref{erfc})
is small, we have $\delta_{z_f}-\deltac\propto \sigma(f\Mh)$.
 For galactic haloes, $\sigma (f\Mh)\gg 1$ in all the 
cosmogonies considered here, and so
$\delta_{z_f}-\deltac\approx \delta_{z_f}\propto (1+z_f)$.
Writing the {\it rms} mass fluctuation defined in equation
(\ref{sigma}) as $\sigma ^2(r_0)\propto r_0^{-(3+\neff)}$,
we have $(1+z_f)\propto (f\Mh)^{-(3+\neff)/6}$. 
On galactic scales, the effective power index
$\neff\sim -2$ for our CDM models, so that
more massive haloes have lower values of $z_f$, and so 
also of $c$. The value of $\alpha$ is slightly larger
for SCDM, because $\neff$ is slightly larger in this case. 
As shown by Navarro et al (1997), if $n_{\rm eff}$ differs 
substantially from $-2$, then $\alpha$ can be very different from 3. 
The effect of changing the {\it amplitude} of the fluctuation spectrum
is negligible, and the slope is also insensitive to changes in $\md$,
$\jd$, and the formation redshift of disks, provided none of
these parameters varies with $\vcir$.  The observed slope of the 
TF relation is thus a generic prediction of hierarchical models 
with CDM-like fluctuation spectra.

\subsubsection {Tully-Fisher scatter}

The observed scatter in the TF relation is quite small; Giovanelli
et al (1997) quote 0.35 magnitudes as an overall measure of scatter, 
while Willick et al (1995,1996)
find $0.4$ magnitudes to be typical. It is clear from Figure 6 that 
we do indeed predict a small scatter, but it is important to
understand why. In the model of Section 2.2 the gravitational 
effect of disks is neglected, all haloes of the same mass have
identical singular isothermal density profiles, and all disks have the same 
$\fd$ and $\mtold$. With these assumptions the TF relation has no
scatter even though the broad $\lambda$-distribution leads to a wide
range of disk sizes and surface brightnesses.
In the more realistic model of
Section 2.3, scatter in the TF relation can  
arise from two effects. For given disk and halo masses, the 
maximum rotation is larger
for more compact disks, i.e. for those residing in more
slowly spinning haloes. As a result, the scatter in the 
$\lambda$-distribution of haloes translates into a scatter about 
the predicted TF relation. This is the source of scatter
in Figure 6; this scatter remains small because the stability requirement
eliminates the most compact systems. In addition, even 
for given $\Md$, $\Mh$ and $\lambda$, the disk rotation
velocity will be larger in more concentrated haloes. Thus 
any scatter in $c$ will produce scatter
in the TF relation. We now examine the consequences of
these two effects in detail.

The solid curves in Figure 7 show the scatter about the mean
model TF relations as a function
of $\lambda_l$, an assumed lower limit on $\lambda$ for the haloes
allowed to contribute observable disks. For given $\fd$, this function
is sensitive neither to cosmological parameters nor to the redshift
of disk formation. The scatter increases as haloes with 
lower values of $\lambda$ are included, because disk self-gravity
becomes important for such systems. It increases with $\fd$, because
the contribution of a disk to the rotation velocity is then larger at
fixed $\lambda$. It is clear that the predicted scatter can be made 
sufficiently small, provided disks with large $\fd$ and small
$\lambda$ are excluded. The arrows next to each curve in Figure 7 show
the smallest value of $\lambda$ for which disks with the given value of
$\fd$ are stable (here we conservatively require $\epsilon_{\rm m}>0.8$).
It is clear that the predicted scatter is smaller than observed 
if $\md\ga 0.025$. For even smaller $\md$, a larger range of
$\lambda$ is allowed for stable disks, leading to a larger
scatter. This is a consequence of the deviation between the NFW
profile and an isothermal sphere. Since only about 10 percent of
haloes have $\lambda<0.025$, the increased scatter in this case
is due to a relatively small number of compact outliers.
  
The distribution of the halo concentration factor at
fixed mass is poorly known.
In the simulations of NFW, the {\it rms} scatter in $c$ is
$\Delta c/c \approx 25\%$. To demonstrate the effect of
such scatter on the predicted TF relation, the dashed curves in
Figure 7 show the predicted scatter
if we assume a 25\% {\it rms} uncertainty in the
concentration parameter of haloes of a given mass.
(Note: NFW show that the scatter in $c$ is uncorrelated with
$\lambda$.)  If this variability in halo profiles is indeed realistic,
it is clear from Figure 7 that it adds relatively little scatter to the
TF relation. As NFW point out, this is in
conflict with the conclusions of Eisenstein \& Loeb (1996) from
an analytic assessment of the expected variation in halo profiles.
In summary, the analysis of this section suggests that it may not be
difficult for hierarchical models to produce a disk population with
Tully-Fisher scatter within the observed limits. We note, however,
that we have not included the bulge contribution to the luminosity of
the galaxies or possible variations in $\Upsilon_d$. Both presumably act
to increase the TF scatter.

\subsubsection {The Tully-Fisher zero-point}
   
The zero-point of the predicted TF relation depends weakly on 
the distribution of $\lambda$ and on our stability requirement. On the
other hand, it depends strongly on $\md$, the fraction of the halo 
mass in disks, on $\mtold$, the
stellar mass-to-light ratio of the disks, and on $z$, the redshift at
which disk material is assembled within a single dark halo.
Figure 8 shows the values of $\mtold h^{-1}$ required to reproduce
the observed zero-point of the $I$-band TF relation 
(equation \ref{zero-point}) as a function of $\md$. 
Zero-points are calculated for stable disks with 
$\epsilon_{\rm m}>1$ (the results are very similar
for $\epsilon_{\rm m}>0.8$ or $>1.2$). 
In all cases the required mass-to-light ratio
is approximately proportional to $\md$. The three arrows
in each panel show the values of $\md$ 
at which 10\%, 50\% and 90\% of the haloes can host
stable disks.

Stable disk galaxies are common and there are no obvious observational
counterparts for a large number of ``failed'' unstable disks. The $z=0$
lines in this figure thus suggest $\mtold < 1h$ in the $I$-band 
for the CCDM model, whereas for SCDM, $\LCDM$ and $\tCDM$ the upper 
limit gets progressively weaker. This trend is a consequence of the steadily
decreasing concentration of haloes of given mass along this sequence,
which results in a corresponding decrease in the maximum rotation velocity.
The same trend can be seen quite clearly in Figure 6. The maximum allowed 
mass-to-light ratio is even smaller for higher assembly redshifts, 
because disks are then less massive for a given $\vcir$. 
The horizontal dotted lines in Figure 8 bracket the $\pm 1\sigma$ range
derived  by Bottema (1997) from his analysis of the dynamics of
observed disks. This range takes account of the fact that some (dim) disks
may have as much as half of their mass in gas. 
As one can see, in order to reproduce the Tully-Fisher zero-point with
$\Upsilon_I$ in the allowed range, normal disks could form at high 
$z$ only if $\md$ is large, while for low formation redshifts
$\md < 0.05$ is possible. For $\md > 0.05$ most haloes {\it cannot} 
host stable disks, so we infer that normal spiral
disks must form late with an effective assembly redshift
close to zero. A reasonable zero-point then requires
$\md\ga 0.03$, in good agreement with the values obtained 
above from the observed sizes and surface brightnesses of disks. 
Finally we note that because disk self-gravity is less important 
for larger values of $\lambda$, somewhat higher mass-to-light 
ratios are needed if LSB galaxies are to fit on the same Tully-Fisher
relation as normal spirals.

\section {HIGH REDSHIFT DISKS AND DAMPED LYMAN ALPHA SYSTEMS}

Damped Ly$\alpha$ absorption systems (DLS), corresponding to 
absorbing clouds with a neutral column density, $\nhi\ga 2\times
10^{20}\cm^{-2}$, are seen in the spectra of many high redshift
QSOs. Their abundance varies very roughly as ${n}(z)\sim 0.03
(1+z)^{1.5}$ over the redshift range $1.5<z<4$, and their total 
HI content is a few tenths of a percent of the critical density, 
about the same the total mass in stars in
the local universe (see e.g. Storrie-Lombardi et al 1996). 
Over the last decade Wolfe and his collaborators have used a wide
range of observational indicators, most recently the
velocity structure in the associated metal lines, to argue that these
systems correspond to large equilibrium disks with $\vcir\ga
200\kms$,  the extended and gas-rich progenitors of present-day 
spirals (Wolfe et al 1986; Lanzetta, Wolfe \& Turnshek 1995; Wolfe 
1995; Prochaska \& Wolfe 1997, but see Jedamzik \& Prochaska
). Early theoretical studies showed
that the total HI content of these systems can be reproduced in 
some but not all hierarchical cosmologies, provided systems with 
circular velocities well below 200 km/s are allowed to contribute
(Mo \& Miralda-Escud\'e 1994; Kauffmann \& Charlot 1994;
Ma \& Bertschinger 1994; Klypin et al 1995). A detailed study of the
standard CDM cosmology by Kauffmann (1996b) concluded that disks in all
haloes down to $\vcir\sim 50$ km/s must contribute, once a realistic 
model for star formation is included. Numerical simulations by Gardner
et al (1997) led these authors to a similar conclusion, while higher 
resolution 
simulations by Haehnelt, Steinmetz \& Rauch (1997) showed that the 
kinematic data of Prochaska \& Wolfe (1997) can be reproduced by 
nonequilibrium disks in low mass haloes and so do not {\it require}
large $\vcir$. These simulations did not include star formation. 
The cross-sections they imply should be viewed with caution 
since their resolution is quite poor and their underlying physical 
assumptions lead to a substantial underprediction 
of the sizes of present-day disks.

Our models allow a more detailed study of the sizes and
cross-sections of high redshift disks than does that of Kauffmann
(1996b). Figure 9 shows how the predicted abundance of absorbers at $z=2.5$
rises as disks within lower and lower mass haloes are allowed
to contribute. We parametrize the mass of the smallest contributing
halo by $V_l$, the rotation velocity of its central 
disk measured at $3\Rd$ and averaged over the 
$\lambda$-distribution of stable disks. 
This abundance refers to absorbers with $\nhi>N_l$ 
and is calculated from  
\begin{eqnarray} \label{nofz}
n(z)&=&{\pi \over 2}(1+z)^3{dl\over dz}
\int d\lambda p(\lambda)\int^\infty_{N_l} {d\nhi \over \nhi ^3}
\int^\infty_{M_l}d\Mh 
n_h(\Mh,z) \Rd ^2(\Mh,\lambda,z)
\nonumber \\
&\times &
N_0^2(\Mh,\lambda,z)(1+2x_1)e^{-2x_1}\Theta(
\epsilon_{\rm m}-\epsilon_{\rm m, crit}),
\end{eqnarray}
where $l(z)$ is the proper distance at redshift $z$,
$n_h(\Mh,z) d\Mh$ is the comoving number density of galactic
haloes given by equation (\ref{ps_mf}), $N_0$ is the central column density
of the disk, $N_l=2\times 10^{20}\cm^{-2}$ is the minimum HI column  
density for a DLS, and $x_1=-{\rm ln}\,[{\rm min}(1,\nhi/N_0)]$.
The step function, $\Theta(x)=1$ for $x>0$ and $\Theta(x)=0$ for $x\le 0$,
ensures that only stable disks are used. For this plot 
we assume $\md=\jd=0.05$ and $\epsilon_{\rm m, crit}=1$. We also
assume that all baryons at the outer edge of the absorbing region are
in the form of HI gas, and that the column density never drops below
$N_l$ at smaller radii. This may be unrealistic at low redshift because
star formation may substantially reduce the mass of HI gas. 
The horizontal dotted lines show a $\pm 2\sigma$ range for
the observed DLS abundance at $z=2.5$, taken from Storrie-Lombardi et al 
(1996). The lower limit on $V_l$ needed to reproduce the observations
can be read off immediately.  This limit varies quite strongly between
cosmogonies. For $\tCDM$, which has the smallest power on galactic scales, 
one has to include disks with rotation velocities
smaller than $50\kms$. This confirms the result found earlier for 
the MDM model which has a similar power spectrum
(Mo \& Miralda-Escud\'e 1994; Kauffmann \& Charlot 1994;
Ma \& Bertschinger 1994; Klypin et al 1995). 
For the other three models, about one third of the observed 
systems can arise in disks with $\vcir\ga 200\kms$, 
and two thirds or more in disks with $\vcir\ga 100\kms$.
It may not be difficult to accommodate disks with $\vcir$ as large
as suggested by Wolfe and collaborators, although these will be 
predicted to be much smaller at $z=2.5$ than nearby disks with the
same circular velocity.

It is interesting that the total cross-section of 
rotationally supported {\it stable} disks with $\vcir\ga 100\kms$ 
can already explain the observations. Unless
the formation of such disks is prevented, they should give rise to
a large fraction of the observed DLS. Some absorbers
may correspond to {\it unstable} disks, but
the abundance of this population should be relatively small;
such disks are compact and so have small cross-sections.
Figure 10 shows the cross-section weighted distribution of the spin
parameter for disks at $z=2.5$. Comparing this with the unweighted 
$\lambda$-distribution (shown as the dot-dashed curve), we see that
absorption comes primarily from
disks with large $\lambda$. For $\md= \jd=0.05$, 
$\lambda_{\rm crit}\approx 0.05$, and less than 20\% of DLS are produced
by unstable systems (assuming, of course, that instability does not
change the cross-section). The value of
$n(z)$ for stable disks is maximised for $\md\sim 0.05$, but varies
little for $\md$ ($=\jd$) in the range from 0.025 to 0.07. 
The abundance is smaller both for larger and for
smaller $\md$, in the first case because the number of stable disks 
is small, and in the second because the amount of gas is small. 
This range is similar to the one we inferred in earlier sections
from the properties of local disk galaxies. As shown by Haehnelt 
et al (1997) many DLS may be produced by gas which has not fully 
settled into centrifugal equilibrium. We are implicitly assuming 
that the cross-section for damped absorption does not vary
strongly, at least on average, over this settling period.
   
It is also instructive to look at the predicted distribution of 
the impact parameter $b$, defined as the distance between the
line-of-sight and the centre of a disk which produces a DLS. 
This distribution can be written as
\beq \label{pofb}
P(b) db\propto db\int d\lambda p(\lambda)\int^\infty_{M_l}
d\Mh n_h(\Mh,z) F[N_0(\Mh,\lambda,z)/N_l,b,\Rd(\Mh,\lambda,z)]   
\Theta(\epsilon_{\rm m}-1),
\eeq 
where $M_l$ is the lower limit on halo mass, and $F$ is the distribution
of $b$ for a disk with scalelength $\Rd$ and central 
surface density $N_0$ observed at a random inclination angle.     
The function $F$ can be obtained from simple Monte-Carlo simulations.
The solid curves in Figure 11 show $P(b)$ at $z=2.5$, with $M_l$ 
chosen to reproduce the observed abundance of DLS. 
For $\tCDM$, this distribution is narrow and 
peaks at quite small values of $b$; DLS at large impact 
parameters are rare in this model. 
For the other three models, however, the upper quartile
of the distribution is at $\sim 5\kpch$, so
systems with relatively large impact parameter do occur. Not
surprisingly, they tend to be associated with high-$\vcir$ disks, as
can be seen from the dashed curves.
At present, there are very few measurements of the impact parameter
for a high redshift DLS. Djorgovski et al (1996) have identified
a galaxy responsible for a DLS at $z=3.15$, and for this case 
the impact parameter is about $13\kpch$ (for $q_0=0.1$).
An observation by Lu et al (1997) suggests that
this is indeed a rotating disk, with $\vcir \sim 200\kms$.
As one can see from Figure 11, 
for this $\vcir$, the observed impact parameter
is not exceptionally large. One cannot draw any statistical 
conclusions from a single event, but if systems like this turn out to
be common the $\tau$CDM model would be disfavored.

It is worth noting that this consistency with the observational results 
requires that $\md$ be in the range 0.025 to 0.07, and that little angular 
momentum be transferred to the dark halo during disk formation. 
Significant angular momentum loss would produce smaller disks.
Haloes with smaller $\vcir$ would then have to be included
in order to match the observed abundance. In this case,
typical DLS would be predicted to have smaller $\vcir$ and smaller
$b$, but to have larger column densities. Substantial star formation
would be required to reduce the column densities to the observed
values, contradicting, perhaps, the low metal abundances measured in
high redshift DLS.
    
The galaxies responsible for DLS are likely to be easier to 
identify at lower 
redshift. It is therefore interesting to give predictions
for such systems. As an example, we show the 
impact parameter distribution for DLS at $z=1$ in Figure 12.   
Results are shown only for SCDM and $\LCDM$; those for 
CCDM and $\tCDM$ are similar.
The redshift is chosen to be typical for the low-redshift DLS
observed by, for example, Steidel et al (see Steidel 1995). In computing this
distribution we have used only disks with $\vcir=100-300\kms$.
This choice is based on the fact that disks with smaller $\vcir$
are difficult to identify, while those with larger $\vcir$ are rare.
An upper limit on $\vcir$ has to be imposed in our calculation, 
because massive haloes are likely to host elliptical galaxies
or galaxy clusters which contain only a small amount of HI gas. 
The median and the upper quartile of the
distribution are at about 7 and $10\kpch$,
respectively (see the solid curves). Larger impact parameters
are expected for samples biased towards galaxies with larger $\vcir$ 
(see the dashed curves). As before, of course, these predictions
assume that star formation has not substantially reduced the gas
column near the damped absorption boundary. Some decrease in
impact parameter may therefore be expected by $z=1$.
However, the absorption cross-section of an exponential disk depends
only logarithmically on $\md$, and so Figure 12 will not change much
provided star formation is only moderately efficient in the outer
galaxy. On the other hand, the HI mass in a DLS decreases in 
proportion to $\md$, so that systems observed at lower redshift may
have systematically lower column densities and so give rise to a
smaller cosmic density parameter in HI.  These trends
are indeed observed (Wolfe 1995), and arise naturally in models which
take star formation into account (Kauffmann \& Charlot 1994; Kauffmann
1996b).      

To see what kind of disks produce DLS
at low redshifts, we can examine the cross-section weighted
distribution of the spin parameter. This distribution 
depends only weakly on $z$, so that Figure 10 is still relevant.
Notice again the strong bias towards the large $\lambda$ tail.  
As discussed in \S 3, the disks which form in haloes with large 
$\lambda$ have large size and low surface densities. We thus
expect that galaxies selected as DLS will be biased towards
low surface densities. As noted by McGaugh \& de Blok (1996) and
others, the star formation rates in such galaxies are low,
giving rise to low surface brightnesses, and producing a relatively
small reduction in their cross-section for damped absorption.
The impact parameter distribution of Figure 12 may then apply. 
There are now some observations of
galaxies responsible for low redshift DLS. 
Steidel et al (see Steidel 1995) find the typical impact parameter 
for such systems to be 5-$15\kpch$.
There are also indications that these galaxies  
tend to be blue, with low surface brightness (Steidel 1995)
and low chemical abundance (Pettini et al 1995).  
All these points seem to agree well with our expectations.
Obviously, more such data will 
provide stringent constraints on disk formation models of the kind we
propose here.

\section {THE EFFECT OF A CENTRAL BULGE}

So far we have considered galaxies to contain only disk and halo 
components. In reality, most spirals also contain a bulge. For a 
galaxy of Milky Way type or later, the bulge mass is less than 20 
percent that of the disk, so its dynamical effects are small.
For earlier-type spirals, however, the bulge makes up a larger
fraction of the stellar mass, and its effects may be significant. 
In this section we use a simple model to assess how our results
are affected by the presence of a central bulge. 
We assume bulges to be point-like, to have a mass which is
$\mb$ times that of the halo, and to have negligible angular
momentum. Further, we assume either that the specific angular momentum
of the disk is the same as that of its dark halo (so that
$\jd=\md$ as usually assumed above), or that the total specific angular
momentum of the stellar components is equal to that of the halo 
(so that $\jd=\md+\mb$). The second case may be appropriate if
angular momentum transfer to the halo is negligible, if
low angular momentum gas forms the bulge, and if high angular momentum
gas settles into the disk.

Under these assumptions it is straightforward to extend the model of 
Section 2.2 to include the bulge. Equation (\ref{adiabatic}) becomes
\beq
\Mh_f(r) = \Md(r) + \Mh(r_i) (1-\md-\mb)+\Mb,
\eeq
and $V^2_{c,DM}$ in equations (\ref{vc_nfw})
and (\ref{vc_bdm}) is replaced by
$V^2_{c, DM}(r)+V^2_{c,b}(r)$, where 
$V^2_{c,b}(r)=G \Mb/r$ is the contribution to the circular
velocity from the bulge. With these changes, the
procedure of Section 2.2 can be applied as before
to obtain the disk scalelength, $\Rd$, and the rotation 
curve, $\vcir (R)$.

In Figure 13 we show the predicted rotation curves when disk 
and bulge masses are equal. Results are shown for the two cases 
mentioned above, namely $\jd=\md$ and $\jd=\md+\mb=2\md$.
The same total halo mass and concentration are
adopted as in the top left panel of Figure 2.
The rotation curves now diverge at small radius because
of our unrealistic assumption of a point-like bulge.
In the case where $\jd=2\md$ the disk
is substantially more extended, and the value of $\vcir$ at given
radius is slightly lower, than in the case where $\jd=\md$.   

   To see the effect of the central bulge more clearly,
we compare values of $\Rd$ and $\vcir(3\Rd)$ for the two cases,
$\Mb=\Md$ and $\Mb=0$. Figure 14 shows how the relative values 
vary with $\md+\mb$. This figure assumes $\lambda=0.05$ and $c=10$,
but in fact the ratios are quite
insensitive to $\lambda$ and $c$.
As one can see, for $\jd=\md$ the values of 
$\Rd$ and $\vcir$ do not change much even when half of the accreted 
gas is put into the central bulge. On the other hand, for 
$\jd=2\md$, the disk scalelength increases by a factor
of two, and the rotation velocity drops significantly below
that for $\Mb=0$. This case produces similar results to those 
for $\lambda=0.1$ and $\Mb=0$. 

   The results in Figure 14 have some interesting implications.
If disks have the same specific angular momentum as their
dark haloes, and if the luminosity of a galaxy is proportional to 
its total stellar mass, then the zero-point
of the TF relation is almost independent of the mass of
the bulge. Even in the extreme case where $\Md=\Mb$ and 
disks have twice the specific angular momentum of their haloes, 
the disk rotation velocity for given $\Md+\Mb$
is only reduced by about 20 percent provided $\md+\mb\la 0.1$.
If disk and bulge had the same mass-to-light ratio, 
this would change the zero-point of the TF relation
by about $-0.8\mag$, but since the stellar mass-to-light ratio of bulges 
is undoubtedly higher than that of disks, 
the actual change would be
smaller. In practice, most of the galaxies used in applications of
the TF relation have bulge-to-disk ratios much smaller than one, 
so we expect the effects of the bulges to produce rather little 
scatter in the TF relation.


\section {DISCUSSION}

In this paper, we have formulated a simple model for the formation of
disk galaxies. Although many of the observed properties of spirals 
and damped Ly$\alpha$ absorbers seem relatively easy to explain, 
it is important to keep in mind our underlying assumptions. These
leave open a number of difficult physical questions. 
We have assumed that all galaxy haloes have the universal density profile
proposed by NFW. This appears relatively safe since the original NFW
claims have been confirmed by a number of independent N-body simulations
(e.g. Tormen et al 1996; Cole \& Lacey 1996; Huss et al 1997).
More significantly we assume that all disks have masses and angular
momenta which are fixed fractions of those of their haloes. Successful
models require $M_d/M\ga 0.05$ and $J_d/J\approx M_d/M$.
These relations are not produced in a natural way by existing simulations
of hierarchical galaxy formation. While it is
common to find all the gas associated with galactic haloes within
condensed central objects (so that $M_d\propto M$), 
the gas usually loses most of its angular momentum to the dark
matter during galaxy assembly. This results in $J_d/J\ll M_d/M$ 
and produces disks which are too small (Navarro \& Benz 1991; Navarro
\& White 1994; Navarro \& Steinmetz 1997). The resolution of this
problem may be related to the well-known need to introduce strong 
feedback in order to get a viable hierarchical model for galaxy 
formation (e.g. White \& Rees 1978; White \& Frenk 1991; Kauffmann 
et al 1993). Such feedback is not included in the simulations, but is
required in hierarchical models to suppress early star formation in small 
objects, thus ensuring that gas remains at late times 
to form large galaxies and the observed intergalactic medium. 
Disks may then form from gas which remains diffuse and so loses 
little angular momentum during protogalactic collapse. 
Unfortunately, while such effects may solve the angular momentum problem, 
they call into question the proportionality between disk mass and halo mass.
The disk/halo mass ratios required by our models are, in any case,
much less than the observed baryon mass fractions in rich galaxy
clusters, so substantial inefficiency in assembling the galaxies is
necessary in any consistent model. 

There are several interesting theoretical issues that are
not addressed in this paper. As is well known, disk galaxies, especially
LSB galaxies, have lower correlation amplitudes than 
ellipticals (see e.g. Davis \& Geller 1976; Jing, Mo \& B\"orner 1993;
Mo, McGaugh \& Bothun 1994). Since hierarchical clustering predicts
little correlation between the spin parameter of a halo and its
large-scale environment, this observed clustering difference
cannot be considered a consequence of disk galaxies
being associated with haloes with large $\lambda$. 
However, as discussed above, disk galaxies, 
especially LSBs, form late in our model. On the other hand, elliptical
galaxies, and in particular cluster ellipticals, form at relatively
high redshifts in hierarchical models (Kauffmann 1996a).
Since for given mass, haloes at low redshift are less correlated than
those at high redshift (Mo \& White 1996), the morphology 
dependence of clustering can arise as a consequence of the morphology
dependence of formation time. This effect is seen quite clearly in the
detailed semianalytic galaxy formation models of Kauffmann et al
(1993) and Kauffmann, Nusser \& Steinmetz (1997). Local
environmental effects may also lead to morphological segregation;
for example, extended disks may be destroyed in a high density 
environment by tidal effects and gas stripping (Moore et al 1996). 
Detailed modelling is clearly needed for any quantitative comparison 
with observation.

In this paper we have considered the formation of disk galaxies only. 
It is interesting to examine how other populations might fit into 
such a scenario. Earlier semianalytic modelling has shown that many
of the global properties of giant ellipticals can be understood if
they formed by mergers between disk systems which themselves formed 
at early times in overdense ``protocluster'' regions. As we have seen,
high redshift disks should be substantially denser and more compact 
than present-day disks of similar mass. This may go a long way towards
explaining the high densities of observed ellipticals. Some bulges and
ellipticals may also form in dark haloes with low spin where any disk
would be too dense to be stable. Such ``spheroidal'' systems may have
systematically different properties from those which formed by mergers. 
We intend to return to a number of these issues in future papers.

\acknowledgments

We are grateful to S. Courteau for providing the data plotted in Figure 4.  
We thank Luiz da Costa for helpful discussions.
This project is partly supported by
the ``Sonderforschungsbereich 375-95 f\"ur Astro-Teilchenphysik'' der
Deutschen Forschungsgemeinschaft.

\appendix

\section{Cosmogonies}

We assume that the universe is dominated
by cold dark matter (CDM). The initial power spectrum is 
\beq
P(k)\propto {kT^2(k)},
\eeq
with
\beq
T(k)={{\rm ln}(1+2.34q)\over 2.34 q}
\left\lbrack 1+3.89 q+(16.1q)^2+(5.46q)^3+(6.71q)^4
\right\rbrack^{-1/4}
\eeq
and
\beq
q\equiv {k\over \Gamma h {\mpc}^{-1}}
\eeq
(see Bardeen et al 1986).
Following Efstathiou et al (1992), we have introduced a shape parameter, 
$\Gamma$, for the power spectrum. 
The {\rm RMS} mass fluctuation in top-hat windows with
radius $r_0$, $\sigma (r_0)$, is defined by
\beq \label{sigma}
\sigma^2 (r_0)={1\over 2\pi^2}\int_0^\infty {dk\over k}
k^3P(k)W^2(kr_0),
\eeq
where $W(x)$ is the Fourier transform of the top-hat window function.
The power spectrum, $P(k)$, is normalized by specifying 
$\sigma_8\equiv \sigma(8\mpch)$. 

We use the Press-Schechter formalism (Press \& Schechter 1974) 
to calculate the mass function of dark haloes. In this formalism, 
the comoving number density of dark haloes, with mass in the range 
$\Mh\to \Mh+d\Mh$, at redshift $z$ is
\beq \label{ps_mf}
n_h(\Mh,z)d\Mh =
-\left({2\over \pi}\right)^{1/2} {{\overline \rho}_0\over \Mh}
{\delta_z \over \sigma (r_0)}
{d\ln \sigma (r_0) \over d\ln \Mh}
{\rm exp} \left\lbrack -{\delta_z^2 
\over 2 \sigma ^2(r_0) }\right\rbrack 
{ d \Mh \over \Mh},
\eeq
where ${\overline \rho}_0$ is the mean matter density of the
universe at present time;
$\Mh$ is the mass of the halo, related to the initial 
comoving radius $r_0$ of the region from which the halo formed 
(measured in current units) by 
$\Mh={4\pi \over 3} {\overline \rho}_0 r_0^3$, $\delta_z$ 
is the threshold linear overdensity for collapse at redshift $z$.
According to the spherical collapse model,
\beq \label{deltaz}
\delta_z=(1+z)\deltac {g(a_0)\over g(a)},
\eeq
where $\deltac\approx 1.68$ (see NFW); $g(a)$ is the linear 
growth factor at the expansion factor $a$ corresponding 
to the redshift $z$ ($a_0$, the present value of $a$, is taken to be 1):
\beq
g(a)={5\over 2}\Omega\left\lbrack \Omega^{4/7}
-\Omega_\Lambda +(1+\Omega/2)(1+\Omega_\Lambda/70)\right\rbrack ^{-1} 
\eeq
(see Carroll, Press \& Turner 1992), with
\beq
\Omega\equiv \Omega (a)={\omnow\over a+\omnow(1-a)+\ovnow (a^3-a)},  
\eeq
\beq
\Omega_\Lambda\equiv \Omega_\Lambda (a)
={a^3\ovnow\over a+\omnow (1-a)+\ovnow (a^3-a)}.  
\eeq
 To put the NFW density profile in a cosmological context,
we need to calculate the concentration factor $c$ defined
in equation (\ref{concentration}). 
For this density profile, $c$ is related to the
characteristic overdensity, $\delch$, by equation (\ref{delta_null}).
The appropriate value of $c$ depends on halo formation history
and on cosmology. NFW proposed a simple model
for $c$ based on halo formation time. The formation redshift
$z_f$ of a halo identified at $z=0$ with mass 
$\Mh$ is defined as the redshift by which 
half of its mass is in progenitors with mass exceeding $f\Mh$, where
$f<1$ is a constant. According to the formula given by Lacey \& Cole
(1993) based on the Press-Schechter formalism, the halo formation time
is then defined implicitly by
\beq \label{erfc}
{\rm erfc}\left\lbrack {(\delta_{z_f}-\deltac)
\over \sqrt {2[\sigma^2(f\Mh)-\sigma^2(\Mh)]}}\right\rbrack={1\over 2}.
\eeq
NFW found that the characteristic overdensity of a halo 
at $z=0$ is related to its formation redshift
$z_f$ by
\beq \label{delta_nullzf}
\delch(\Mh,f)=C(f)\Omega_0 \left\lbrack 1+z_f(\Mh,f)\right\rbrack ^3,
\eeq
where the normalization $C(f)$ depends on $f$. We will take $f=0.01$
as suggested by the N-body results of NFW. 
In this case $C(f)\approx 3\times 10^3$. 
Thus, for a halo of given mass at $z=0$, one can obtain the
concentration factor $c$ from 
equations (\ref{deltaz})-(\ref{delta_nullzf}). 
In practice, we first solve $z_f$ from equation (\ref{erfc}) and insert 
the value of $z_f$ into equation (\ref{delta_nullzf}) to get $\delch$,
we then use this value of $\delch$ in equation 
(\ref{delta_null}) to solve for $c$. The distribution of $c$ 
is not well known. In the simulations of NFW, the scatter in $c$ is
$\Delta c/c \approx 0.25$.   

Equations (\ref{erfc})-(\ref{delta_nullzf}) can be easily extended to
calculate concentrations for
haloes identified at $z>0$. In this case, we just replace
$\deltac$ in equation (\ref{erfc}) by $\delta_z$, and 
$\omnow$ in equation (\ref{delta_nullzf}) 
by $\Omega (a)/(1+z)^3$, where $a$ is the scale
factor at $z$.

\begin{figure}
\epsscale{1.0}
\plotone{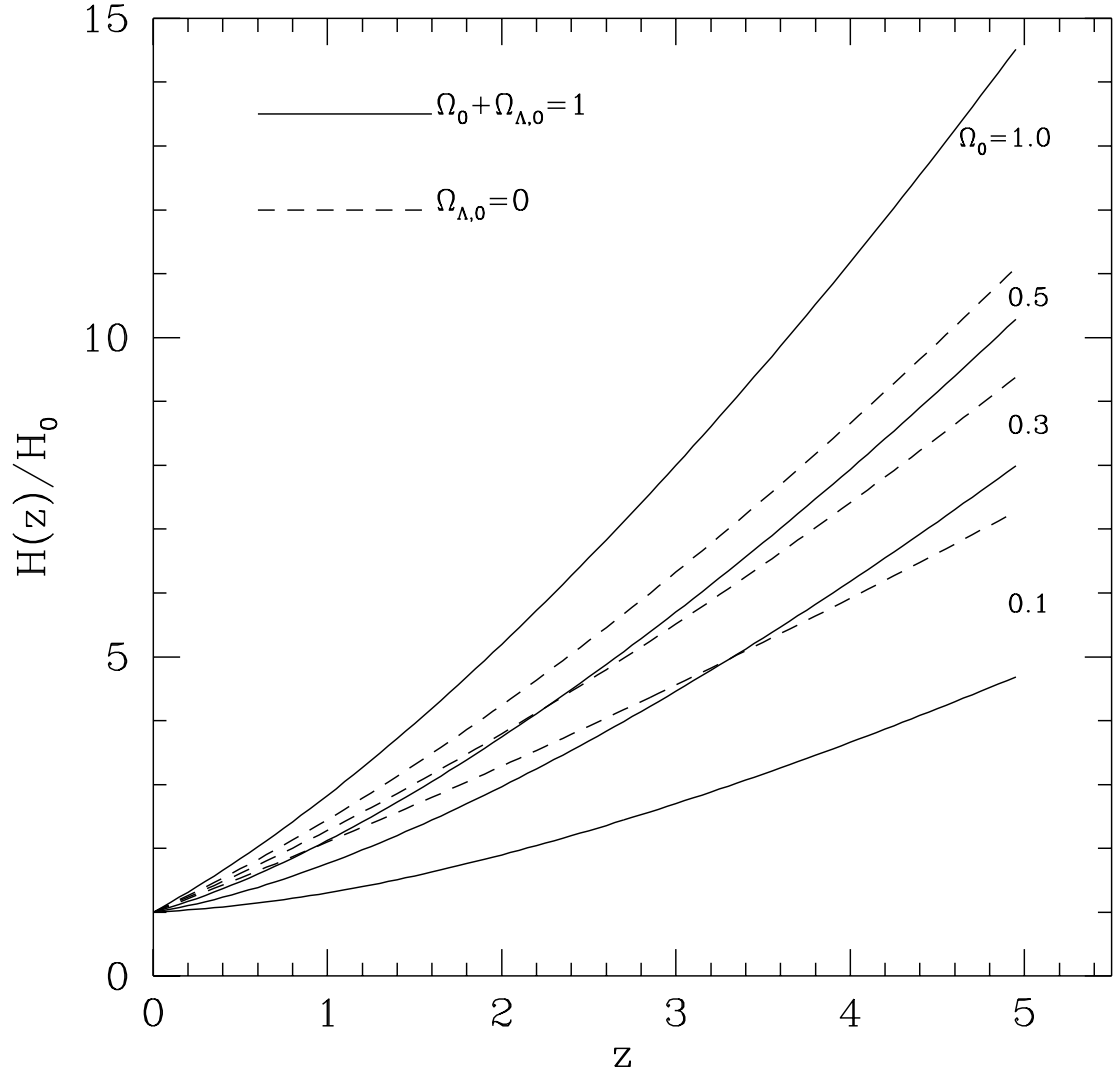}
\caption{The Hubble constant (in units of its present value) as
a function of redshift for flat ($\omnow+\ovnow=1$)
and open models with various $\omnow$. 
}\end{figure}

\begin{figure}
\epsscale{0.75}
\plotone{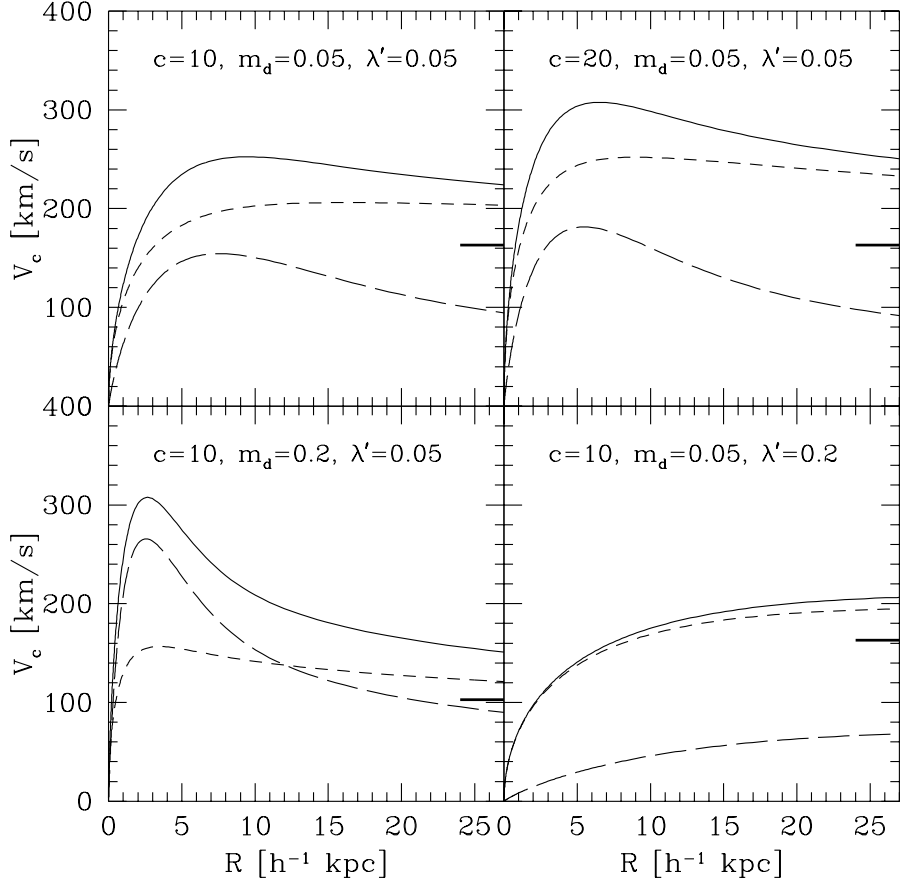}
\caption{
Rotation curves for disk galaxies formed in haloes with an initial 
density distribution described by the NFW profile. The shape
of the rotation curves depends only on the concentration parameter,
$c$, the fraction of mass in the disk, $\fd$, and the spin parameter,
$\lambda^\prime \equiv (j_d/m_d)\lambda $. The total mass of a halo determines both the disk
scalelength and the amplitude of the rotation curve. Four panels are
shown for different sets of $c$, $\fd$ and $\lambda^\prime$. 
The disk mass is assumed to be $M_d=5\times 10^{10} h^{-1}M_\odot$.
The rotation velocities induced by the disk and the dark matter
are shown using long-dashed and short-dashed lines, respectively,
while the total rotation velocity is shown by a solid line.
The rotation velocity induced by the disk
reaches a peak at about two disk scalelengths.
The rotation velocity at $r_{200}$ is indicated by thick horizontal
bars and is identical to the rotation velocity at all radii for a
singular isothermal sphere with the same mass.
The upper left panel should be viewed as the reference
panel. Only one parameter is varied between this panel and any other.
Notice how the prominence and amplitude of the peak in the rotation curve 
change between panels.
}
\end{figure}

\begin{figure}
\epsscale{1.0}
\plotone{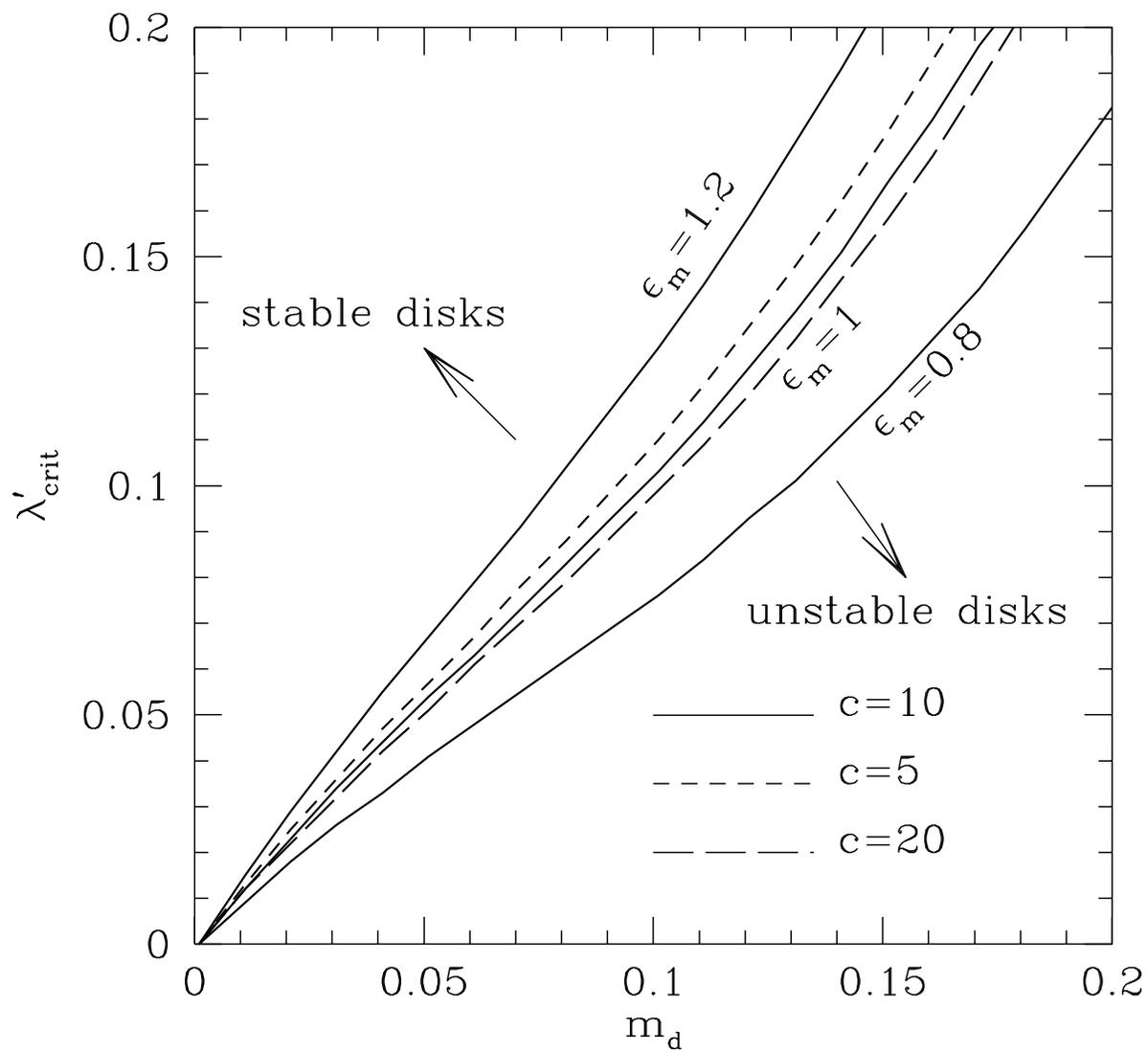}
\caption{Critical values of $\lambda^\prime$ ($\equiv \lambda j_d/m_d$)
for disk instability
as a function of $\md$. Results are shown for three choices
of $\epsilon_{\rm m}$.
For a given $\md$, disks are stable
if $\lambda^\prime>\lambda^\prime_{\rm crit}$. The dependence on halo concentration,
$c$, is weak, and is shown only for $\epsilon_{\rm m}=1$.
}
\end{figure}

\begin{figure}
\epsscale{1.0}
\plotone{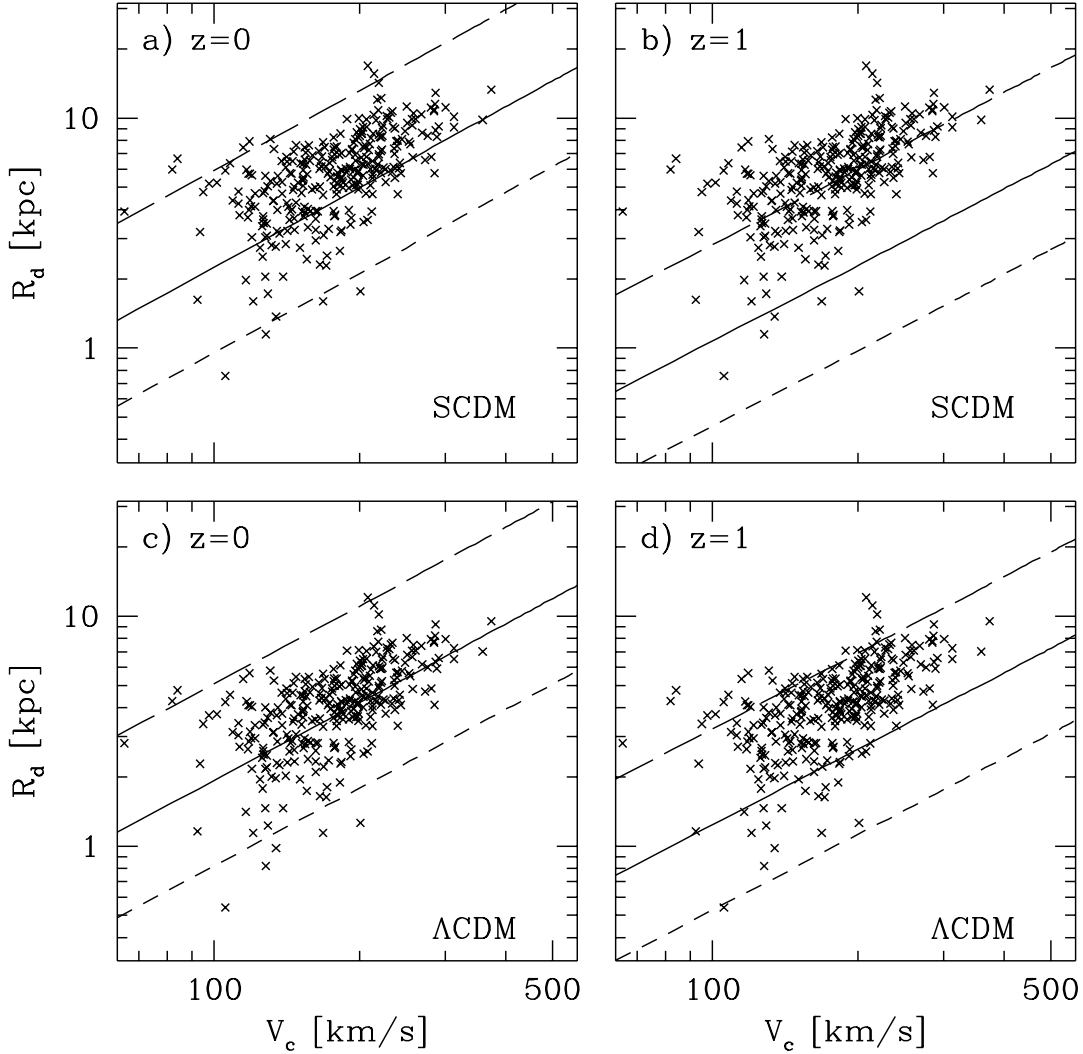}
\caption
{Model predictions for $\Rd$ as a function of $\vcir$ 
for stable disks assembled at  $z=0$ and at $z=1$ in
the SCDM and $\LCDM$ models. The solid lines give the relations for
critical disks when $\md=0.05$, while short-dashed lines give the
corresponding relations for $\md=0.025$. Stable disks must lie above
the line for the relevant value of $\md$. The long-dashed 
lines correspond to $\md=\jd$ and $\lambda=0.1$; at most 10\% of disks
should lie above these lines. The data points are the 
observational results of Courteau (1996; 1997) for a sample of nearby
normal spirals. 
}\end{figure}

\begin{figure}
\epsscale{1.0}
\plotone{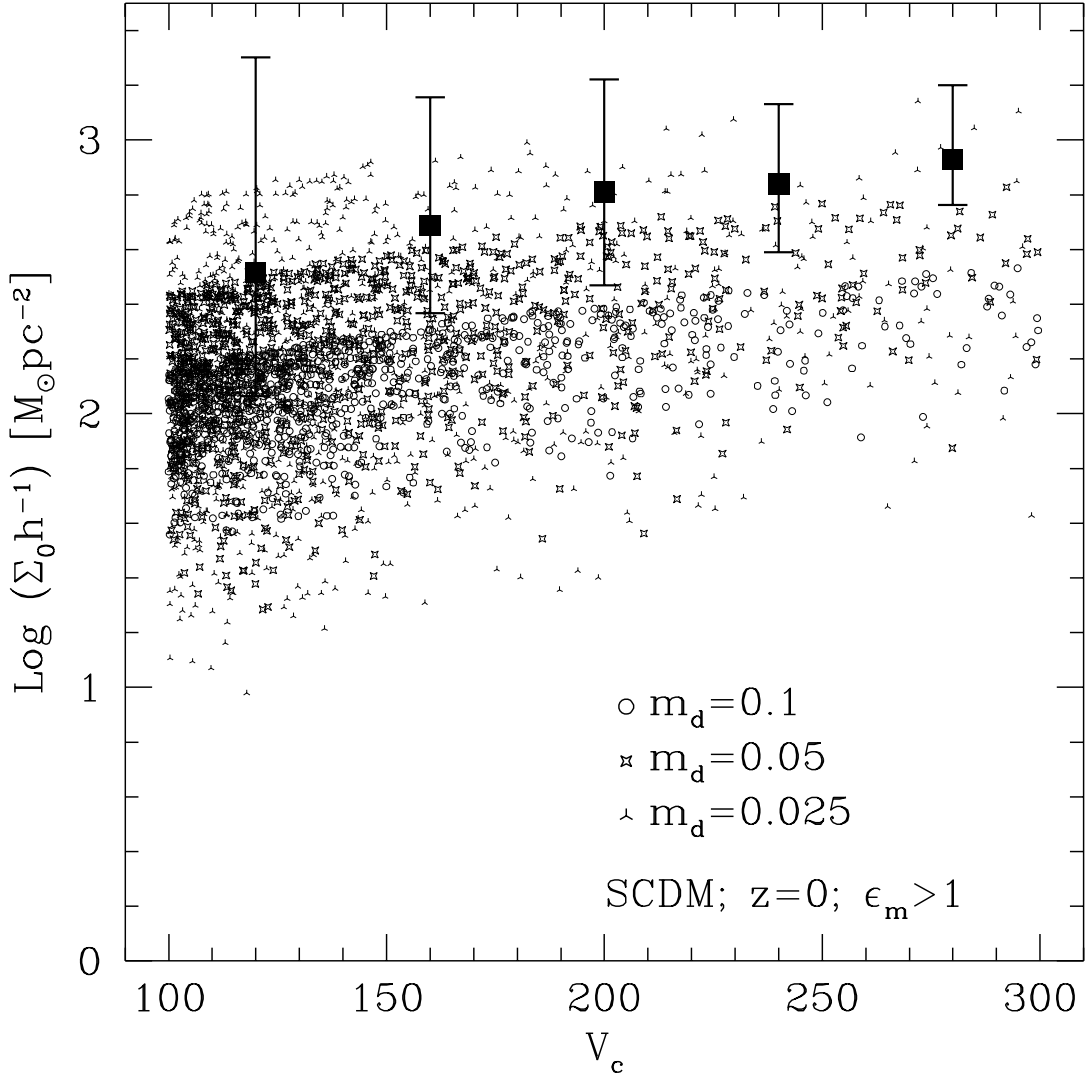}
\caption
{The distribution of central surface density for disks
at $z=0$ in the SCDM model. Results are shown assuming
$\md=\jd$ and for $\md=0.025$, 0.05 and 0.1. Only stable disks
with $\epsilon_{\rm m}>1$ are plotted. The squares with error bars
refer to the observational data from Figure 4. The circular velocities
have been converted to a luminosity using the Tully-Fisher relation of
Giovanelli et al (1997), and this in turn has been
converted into a mass using $\Upsilon_I=1.7h$ (Bottema 1997).
The square is plotted at the median
value in each bin and the ends of the error bars mark the upper and
lower 10\% points of the distributions.
}\end{figure}

\begin{figure}
\epsscale{1.0}
\plotone{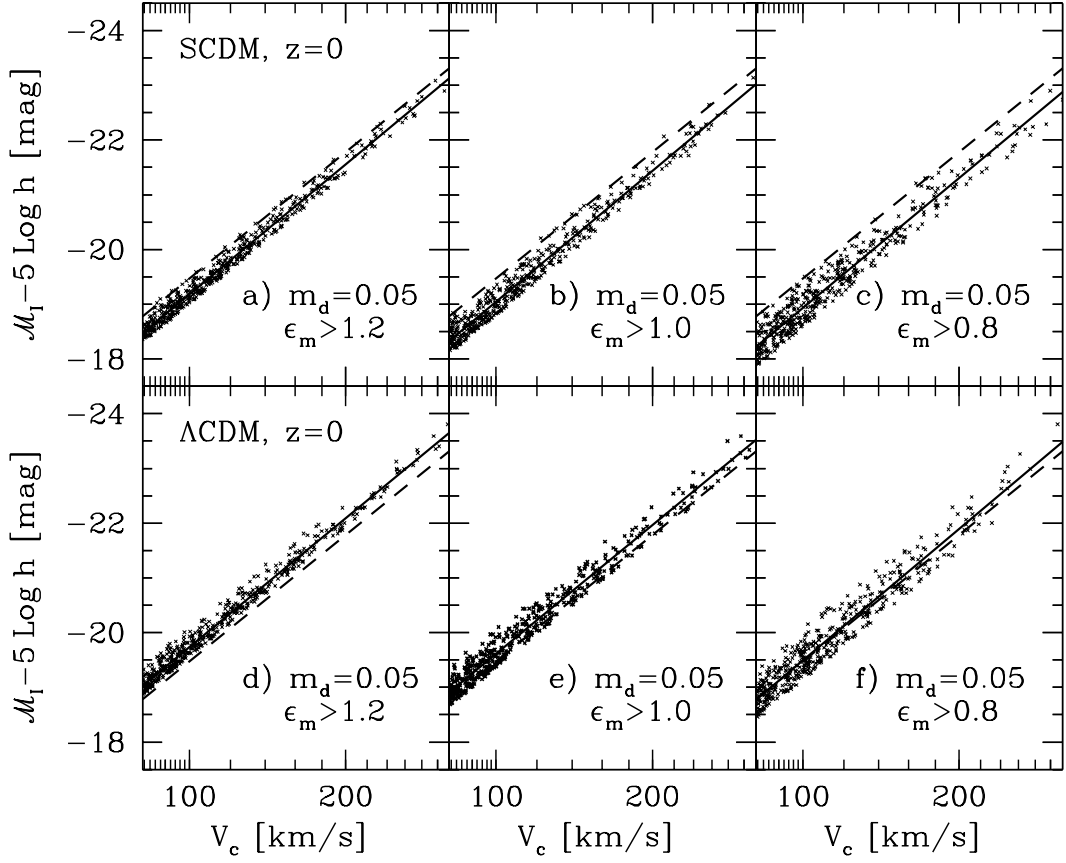}
\caption{Tully-Fisher (TF) relations for stable disks at $z=0$
in the SCDM and $\LCDM$ cosmogonies.
Monte Carlo samples of the predicted luminosity-rotation velocity
distribution are shown for three choices of $\epsilon_{\rm m}$.
We have converted stellar mass (in the model) into I-band luminosity using
$\Upsilon = 1.7h$ (Bottema 1997).
The solid lines give the linear regressions of absolute magnitude
against $\log \vcir$. The dashed lines show the observed TF relation
as given by Giovanelli et al (1997) 
}\end{figure}

\begin{figure}
\epsscale{1.0}
\plotone{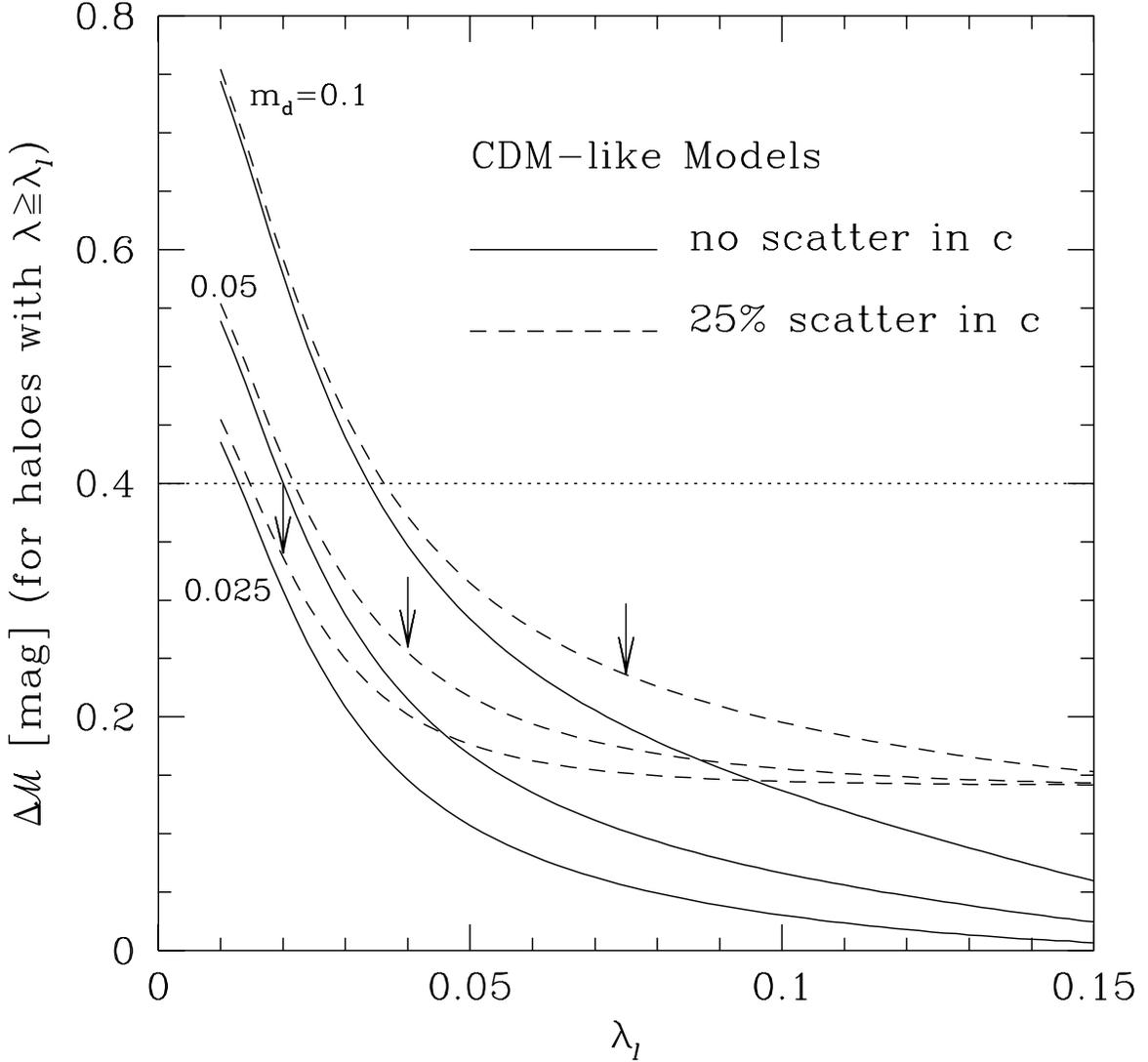}
\caption{
The scatter about the mean TF relation as a function
of $\lambda_l$, the lower limit on $\lambda$ for the haloes
included in the relation.
The solid curves show results which ignore the scatter
caused by variations of the halo concentration, $c$,
while the dashed curves show assume an {\it rms} scatter
of 25\% in $c$. Three curves are shown for three different disk mass 
fractions, $m_d=0.1, 0.05$ and 0.025 (from top to bottom).
The horizontal line shows the observed scatter for
normal disk galaxies, as given by Willick et al (1996).
The arrows mark the critical values of $\lambda$; 
disks are stable only for larger $\lambda$ values.  
}\end{figure}

\begin{figure}
\epsscale{1.0}
\plotone{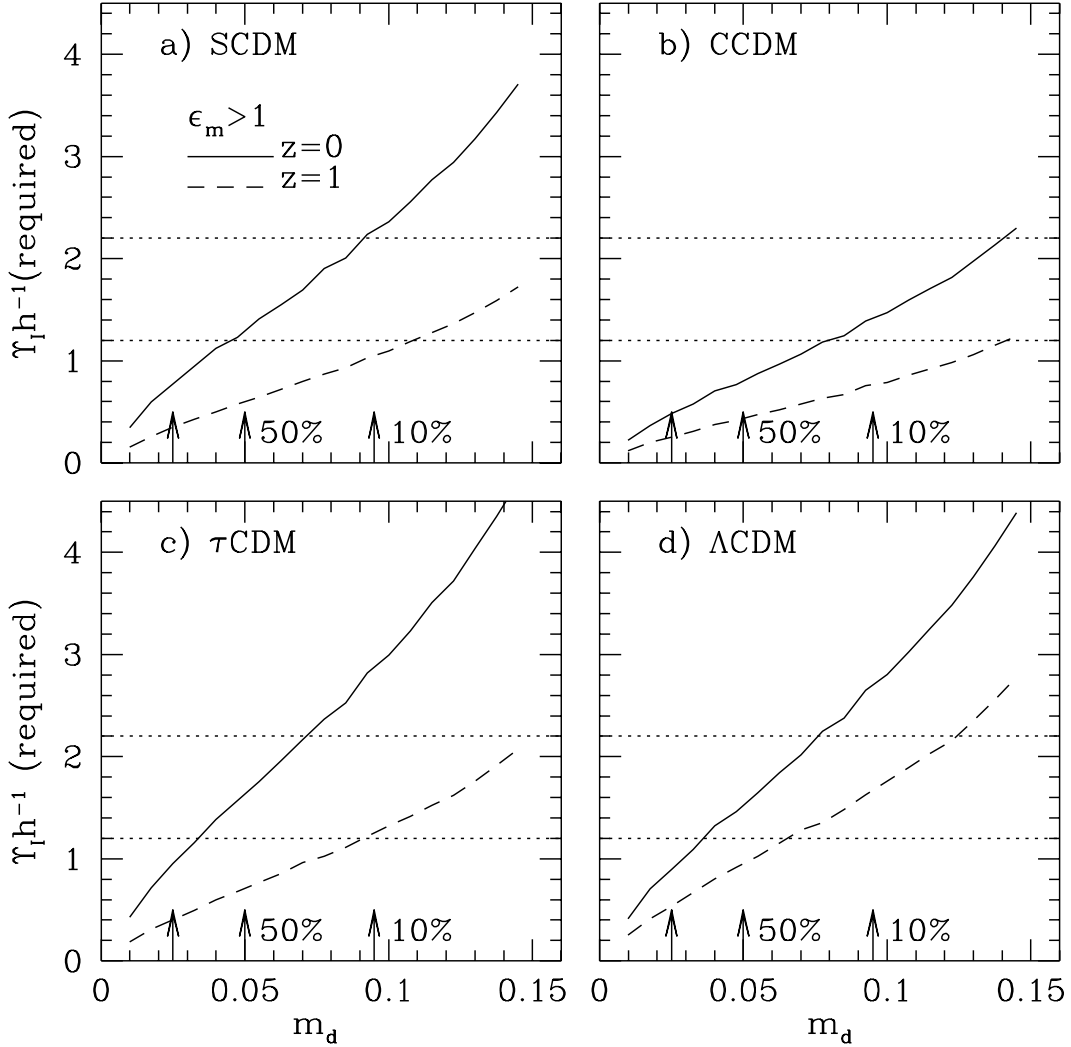}
\caption{The mass-to-light ratio required to reproduce
the observed zero-point of the TF relation as given by Giovanelli
et al (1997). Stable disks with $\epsilon_{\rm m}>1$ are used.
Results are shown for two assembly redshifts, $z=0$ (solid lines) and
$z=1$ (dashed lines). The horizontal lines bracket the values of $\Upsilon_I$
allowed by dynamical studies of disks (Bottema 1997).
The arrows indicate the values of $\md$ at which 90\%,
50\% and 10\% of haloes can host stable disks.   
}\end{figure}

\begin{figure}
\epsscale{1.0}
\plotone{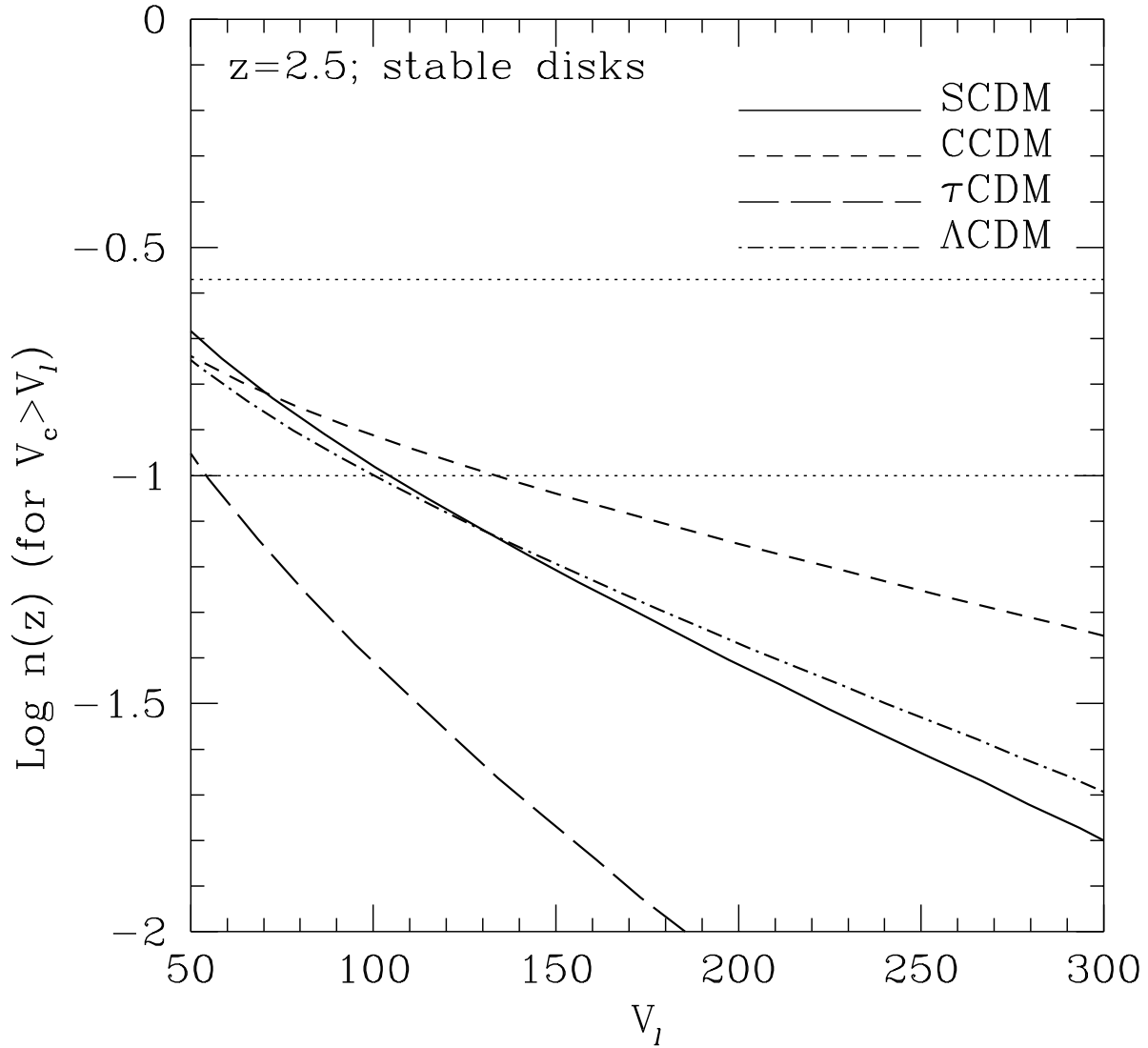}
\caption{
Logarithm of the predicted abundance of damped Ly$\alpha$ systems at $z=2.5$  
as a function of $V_l$, the rotation velocity of the least
massive disks allowed to contribute. Only stable disks are assumed to
give rise to damped Ly$\alpha$ systems (i.e. disks with $\epsilon_{\rm m}>1$). 
The horizontal lines show the $\pm 2\sigma$ range
of the observed abundance as given by Storrie-Lombardi et al (1996).
}\end{figure}

\begin{figure}
\epsscale{1.0}
\plotone{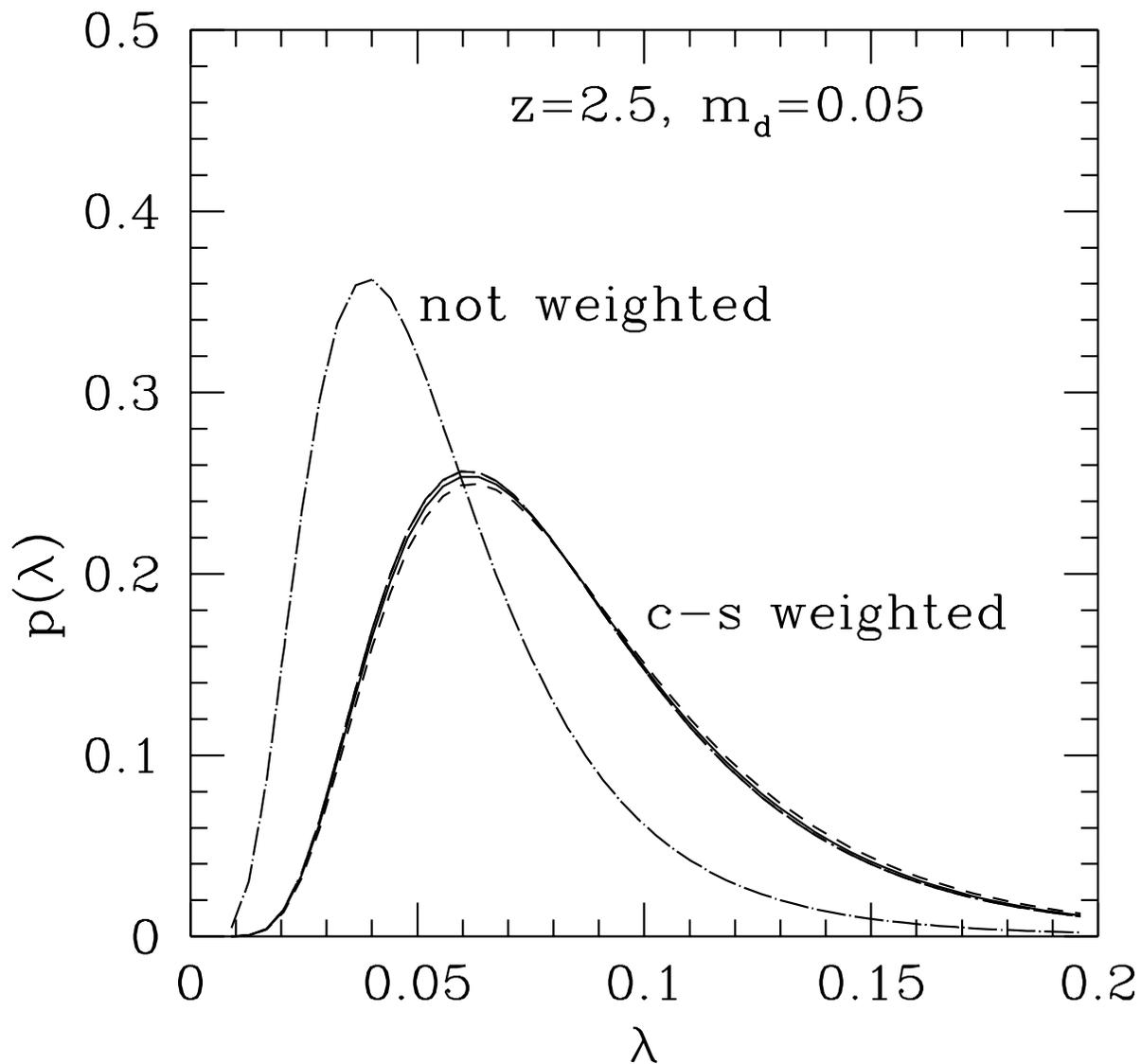}
\caption{
The cross-section (c-s) weighted distribution of halo spin
at $z=2.5$. Notice that this distribution depends only weakly 
on cosmogony; the curves corresponding to our four cosmogonies can
barely be distinguished. The unweighted $\lambda$-distribution, as
given by equation (\ref{plambda}), is shown as the dot-dashed curve. 
}\end{figure}

\begin{figure}
\epsscale{1.0}
\plotone{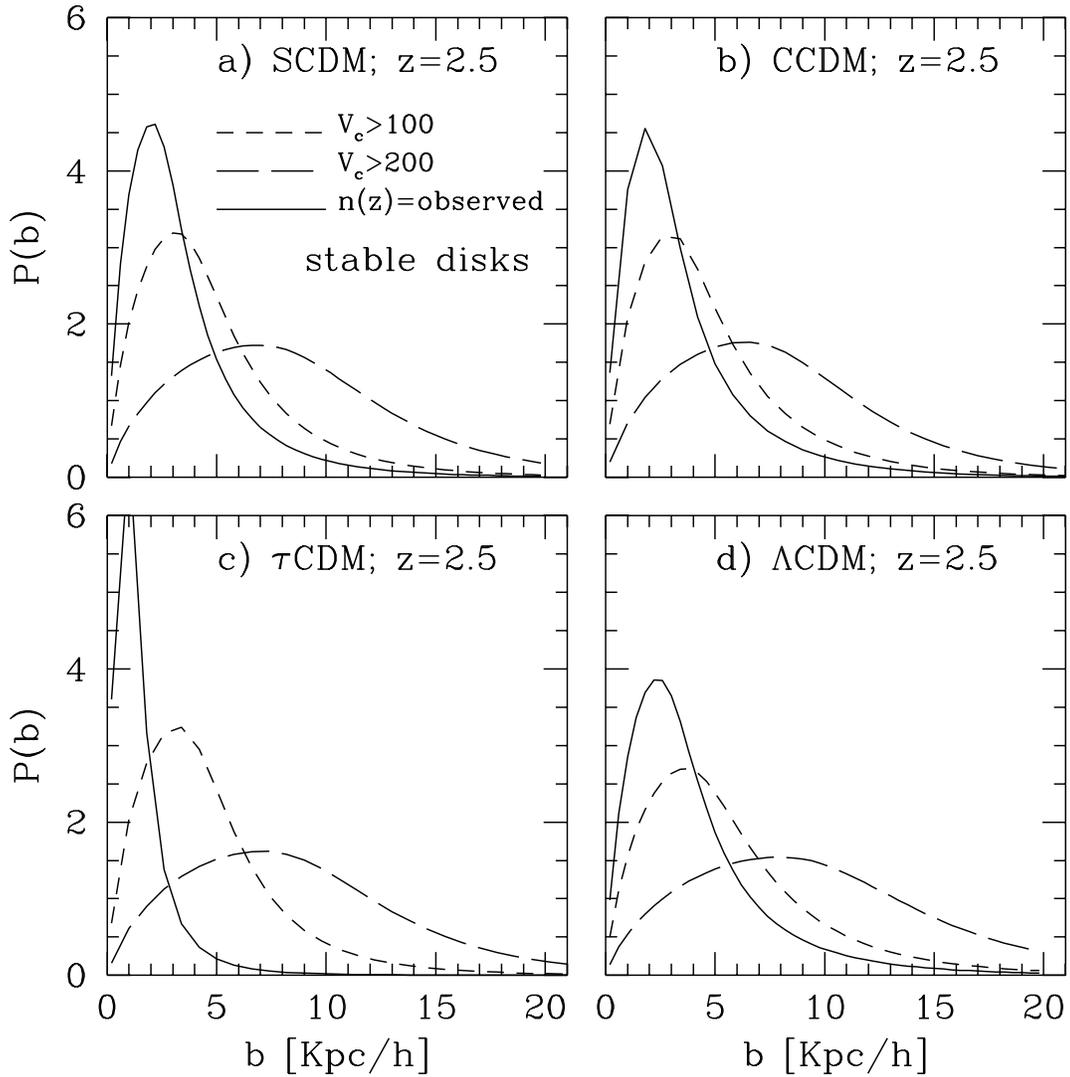}
\caption{
The distribution of impact parameter for
damped systems in stable disks at $z=2.5$. 
The solid curves show the results 
when the mass of the smallest contributing disks 
is chosen so that the predicted abundance of absorbers
is equal to the observed value. 
The dashed curves show the results when only disks with
large $\vcir$ are selected.  
}\end{figure}

\begin{figure}
\epsscale{1.0}
\plotone{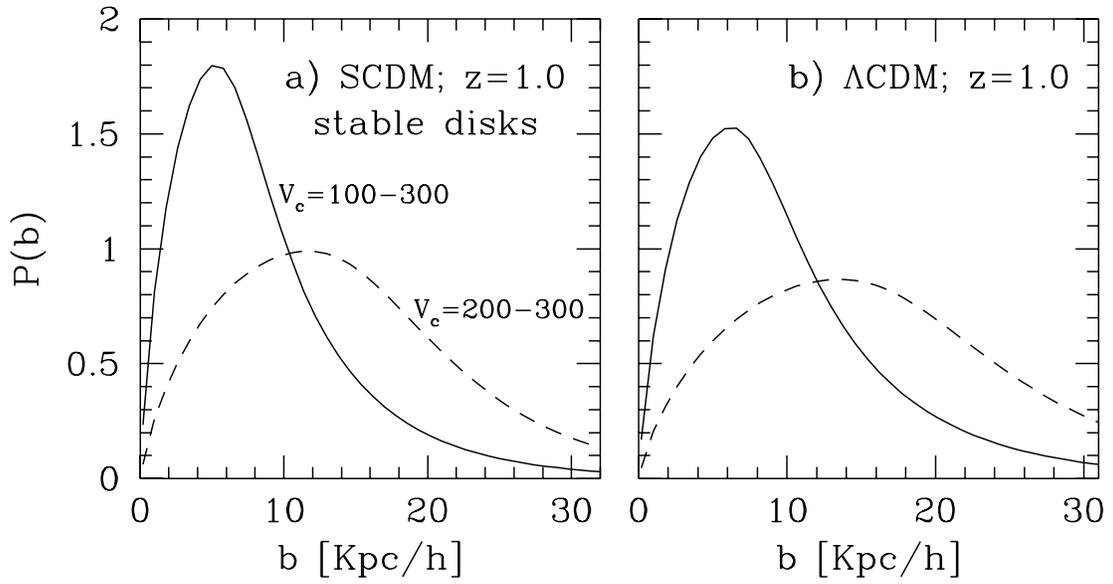}
\caption{
The distribution of impact parameter for
damped systems in stable disks at $z=1$. 
Results are shown for two ranges of $\vcir$.  
}\end{figure}

\begin{figure}
\epsscale{1.0}
\plotone{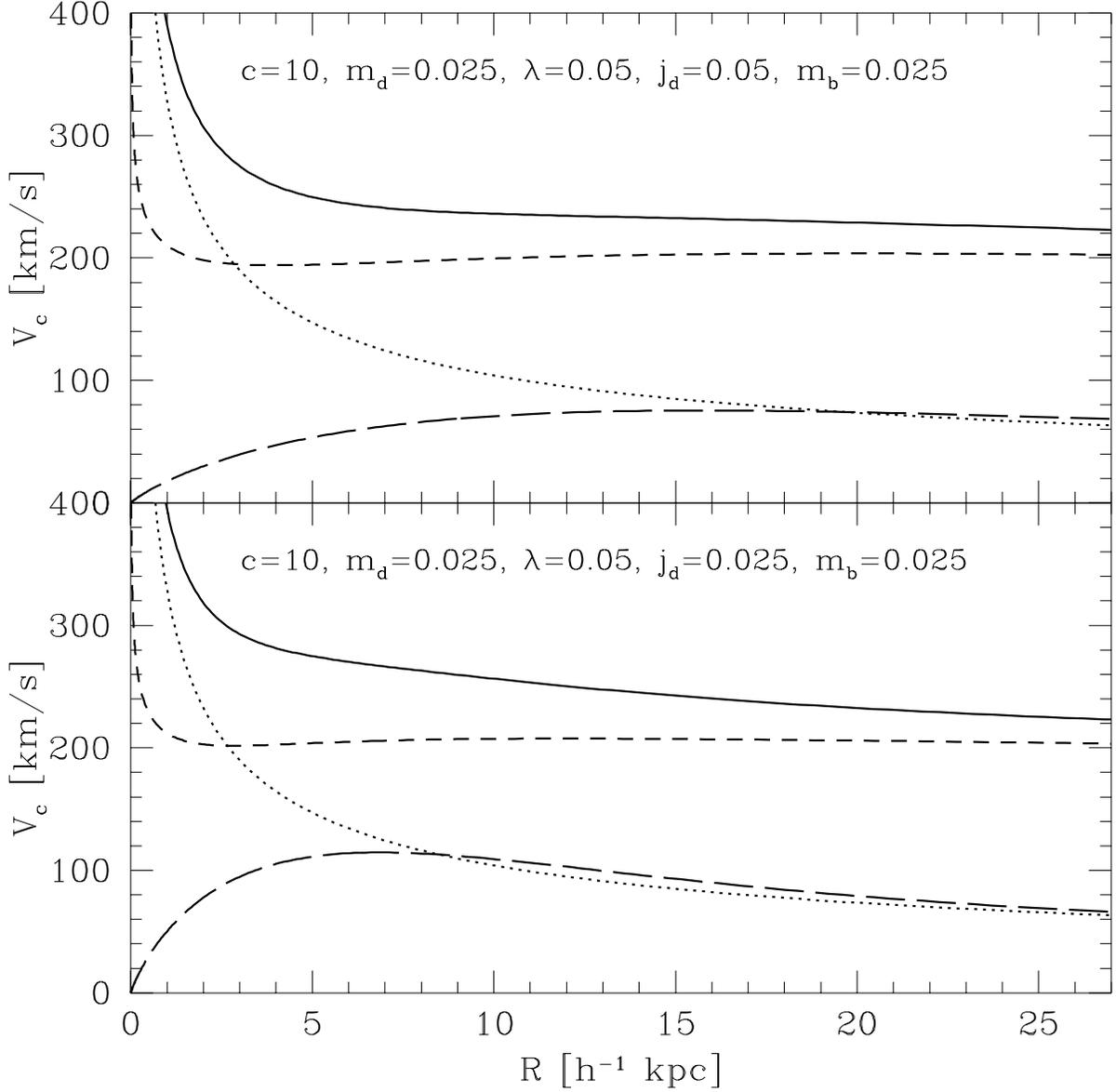}
\caption{
Rotation curves for disk galaxies with central bulges
(cf. Figure 2). The disk and the bulge are assumed to have the same mass,
$2.5\times 10^{10} h^{-1} M_\odot$.
The circular velocities induced by the disk, the bulge and the dark matter
are shown by long-dashed, dotted and short-dashed lines, respectively.
The total circular velocity is shown by the solid line. 
In the top panel, the disk is assumed to have twice as much specific 
angular momentum as the dark matter, while in the bottom panel the two
specific angular momenta are assumed to be the same. The central cusps in
these rotation curves reflect our unrealistic assumption of a
point-mass bulge.
}
\end{figure}

\begin{figure}
\epsscale{1.0}
\plotone{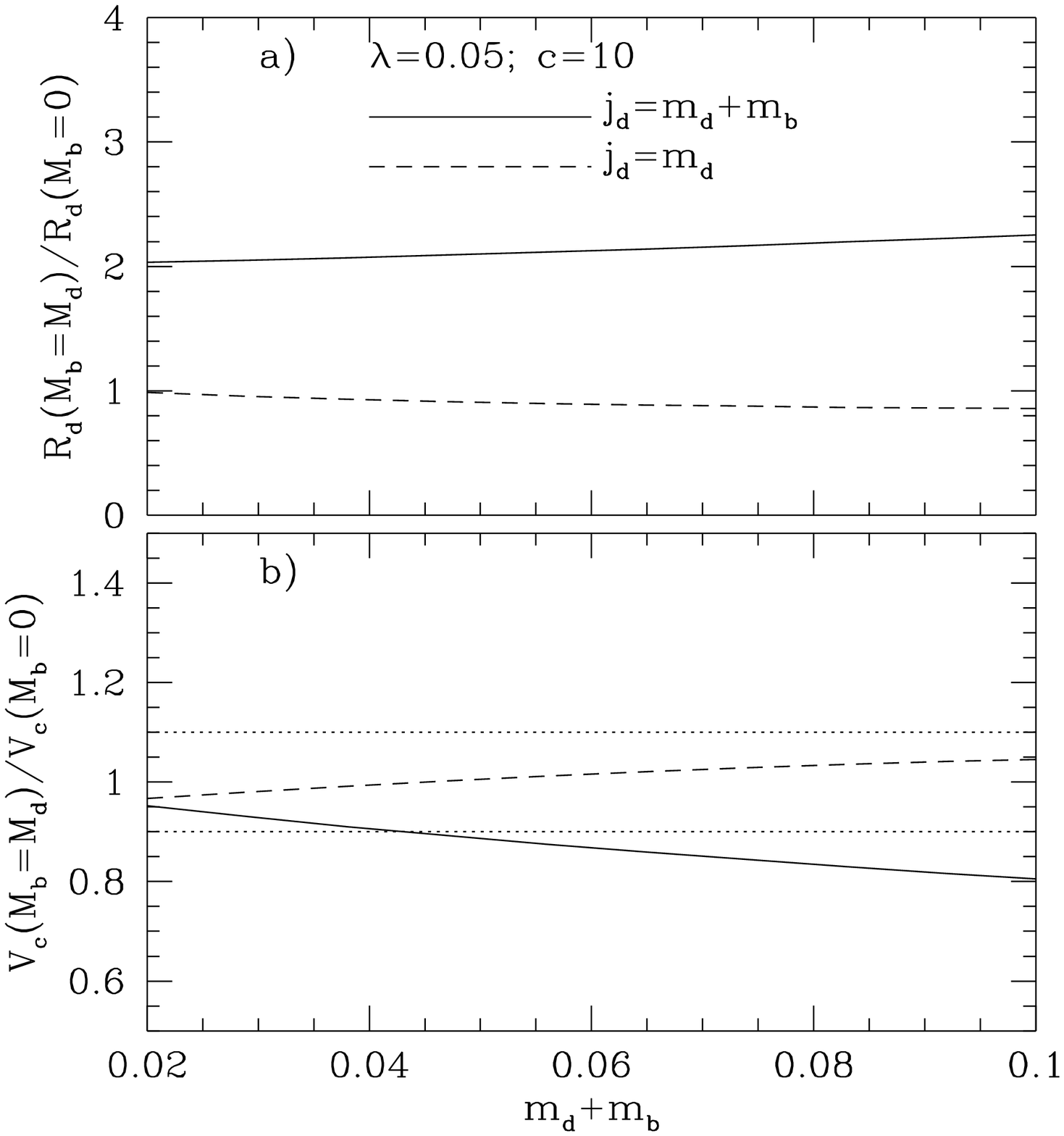}
\caption{
Disk scalelength, $\Rd$,  and disk rotation velocity,
$\vcir(3\Rd)$, in a model where the mass
of the central bulge ($\Mb$) is equal to that of the
disk ($\Md$), as functions of the total stellar mass fraction,
$\md +\mb$. Results are shown in units of the values when $\Mb$ 
is set to zero, We have assumed $\lambda=0.05$ and $c=10$.
The solid curves show the results for $\jd=\md+\mb$, 
while the dashed curves show those for $\jd=\md$.
The two dotted lines give a rough indication of the
scatter in the observed TF relation. 
}
\end{figure}


\newpage

\begin{references}
\reference{} Bardeen J., Bond J.R., Kaiser N., Szalay A.S., 1986,
             {ApJ}, 304, 15 
\reference{} Barnes J., White S.D.M. 1984, MNRAS, 211, 753
\reference{} Baugh C.M., Cole S., Frenk C.S., Lacey C.G. 1997, preprint
 	(astro-ph/9703111) 
\reference{} Bertschinger, E. 1985, ApJS, 58, 39
\reference{} Bessel M.S., 1979, PASP, 91, 589
\reference{} Binney J., Tremaine S., 1987, Galactic Dynamics.
             Princeton Univ. Press, Princeton, NJ
\reference{} Bothun G., Impey C., McGaugh S., 1997, preprint
             (astro-ph/9702156)
\reference{} Bottema, R. 1997, preprint (astro-ph/9706230)
\reference{} Broeils A.H., Courteau S., 1996, preprint (astro-ph/9610264)
\reference{} Carroll S.M., Press W.H., Turner E.L., 1992, ARAA, 30, 499
\reference{} Casertano S. Van Gorkom J., 1991, AJ, 101, 1231
\reference{} Christodoulou D.M., Shlosman I., Tohline J.E., 1995,
             ApJ, 443, 551
\reference{} Cole S., Aragon-Salamanca A., Frenk C.S., Navarro J.F., 
             Zepf S. E., 1994, MNRAS, 271, 781
\reference{} Cole S., Lacey C., 1996, preprint (astro-ph/9510147 v3)
\reference{} Courteau S., 1996, ApJS, 103, 363
\reference{} Courteau S., 1997, ApJS, 1in press
\reference{} Dalcanton J.J., Spergel D.N., Summers, F.J., 1997,
             ApJ, 482, 659 (DSS)
\reference{} Davis M., Geller M., 1976, ApJ, 208, 13
\reference{} Djorgovski S.G., Pahre M.A., Bechtold J., Elston R.,
             1996, Nature, 382, 234
\reference{} Efstathiou G., Bond J.R., White S.D.M., 1992, MNRAS, 258, 1P
\reference{} Efstathiou G., Lake, G., Negroponte, J., 1982, MNRAS,
             199, 1069
\reference{} Eisenstein, D.J., Loeb, A., 1996, ApJ, 459, 432
\reference{} Fall S.M., Efstathiou G., 1980, MNRAS, 193, 189
\reference{} Gardner J.P., Katz N., Weinberg D.H., Hernquist L.,
             1997, preprint (astro-ph/9705118)     
\reference{} Giovanelli R., Haynes M.P., da Costa L.N., Freudling W.,
             Salzer J.J., Wegner G., 1997, preprint (astro-ph/9612072)
\reference{} Gunn J., Gott R., 1972, ApJ, 176, 1
\reference{} Haehnelt M. G., Steinmetz M., Rauch, M., 1997, preprint
	(astro-ph/9706201)
\reference{} Huss A., Jain B., Steinmetz M., 1997, preprint (astro-ph/9703014)
\reference{} Jedamzik, K., Prochaska, J.X. 1997, preprint (astro-ph/9706290)
\reference{} Jing Y.P., Mo H.J. \& B\"orner G., 1991, A\&A, 252, 449.
\reference{} Katz N., Gunn J., 1991, ApJ, 377, 365
\reference{} Kauffmann G., Charlot S., 1994, ApJ, 430, L97
\reference{} Kauffmann G., White S.D.M., Guiderdoni B., 1994, 
             MNRAS, 267, 981
\reference{} Kauffmann G. 1996a, MNRAS, 281, 487
\reference{} Kauffmann G. 1996b, MNRAS, 281, 475
\reference{} Kauffmann G., Nusser A., Steinmetz M., 1997, MNRAS, 286, 795
\reference{} Klypin A., Borgani S., Holtzman J., Primack. J., 1995,
              ApJ, 444, 1
\reference{} Lacey C., Cole S., 1993, MNRAS, 262, 627
\reference{} Lacey C., Cole S., 1995, preprint (astro-ph/9510147)
\reference{} Lanzetta K.M., Wolfe A. M., Turnshek D. A., 1995, ApJ, 440,
	435L
\reference{} Lu L., Sargent W.L.W., Barlow T.A., 1997,
             preprint (astro-ph/9701116)
\reference{} Ma C.P., Bertschinger E., 1994, ApJ, 434, L5
\reference{} McGaugh S.S., 1994, Nature, 367, 538
\reference{} McGaugh S.S., de Blok W.J.G., 1996, preprint (astro-ph/9612070)
\reference{} Mo H.J., Jing Y.P.,  White S.D.M., 1996, MNRAS, 282, 1096
\reference{} Mo H.J., McGaugh S.S., Bothun G.D., 1994, MNRAS, 267, 129
\reference{} Mo H.J., Miralda-Escud\'e J., 1994, ApJ, 430, L25
\reference{} Mo H.J., White S.D.M., 1996, MNRAS, 282, 347
\reference{} Moore B., Katz N., Lake G., Dressler A., Oemler, A. Jr.,
	1996, Nature, 379, 613
\reference{} Navarro J.F., Benz W. 1991, ApJ, 380, 320
\reference{} Navarro J.F., White S.D.M., 1994, MNRAS, 267, 401
\reference{} Navarro J.F., Frenk C. S., White S.D.M., 1995, 
             MNRAS, 275, 720
\reference{} Navarro J.F., Frenk, C. S., White, S.D.M., 1996, ApJ, 462, 563
\reference{} Navarro J.F., Frenk, C. S., White, S.D.M., 1997, in press
	 (preprint, astro-ph/9611107) (NFW)
\reference{} Navarro J.F., Steinmetz M., 1997, 478, 13
\reference{} Pettini M., King D.L., Smith L.J., Hunstead R.W., 1995
            in QSO Absorption Lines, ed. G. Meylan, Springer: Berlin, p.71
\reference{} Press W.H., Schechter P., 1974, ApJ, 187, 425 (PS)
\reference{} Prochaska J.X., Wolfe A.M., 1997, ApJ, submitted
\reference{} Percic, M., Salucci, P., 1991, ApJ, 285, 307
\reference{} Shanks T., 1997, preprint (astro-ph/9702148)
\reference{} Sprayberry D., Bernstein G.M., Impey C.D., Bothun G.D.,
             1995, ApJ, 438, 72
\reference{} Steidel C.C., 1995, 
             in QSO Absorption Lines, ed. G. Meylan, Springer: Berlin, p.137
\reference{} Steinmetz M., Bartelmann M., 1995, MNRAS, 272, 570
\reference{} Steinmetz M., M\"uller E., 1995, MNRAS, 276, 549
\reference{} Storrie L.J., McMahon R.G., Irwin M.J., 1996, MNRAS, 283,
             L79
\reference{} Strauss, M., Willick, J.A., 1995, Phys. Reports, 261, 271
\reference{} Toomre A., 1964, ApJ, 139, 1217
\reference{} Tormen G., Bouchet F.R., White S.D.M., 1997, MNRAS, 286, 865
\reference{} Turner, M.S. 1996, preprint (astro-ph/9610158)
\reference{} Walker T. P., Steigman, G., Schramm, D. N., Olive, K. A.,
              Kang, H.-S. 1991, ApJ 376, 51
\reference{} Warren M.S., Quinn P.J., Salmon J.K., Zurek W.H., 1992,
             ApJ, 399, 405
\reference{} White M., Scott D., 1996, Comments on Astrophysics, 8,
              No. 5
\reference{} White S.D.M., 1995, Les Houches 
\reference{} White S.D.M., Efstathiou G., Frenk C., 1993, MNRAS, 262, 1023 
\reference{} White S.D.M., Frenk C.S., 1991, ApJ, 379, 52 
\reference{} White S.D.M., Rees M.J., 1978, MNRAS, 183, 341
\reference{} Williams R.E., et al, 1996, AJ, 112, 1335
\reference{} Willick J.A., Courteau S., Faber S.M., Burstein D., 
             Dekel A., Kolatt T., 1996, ApJ, 457, 460
\reference{} Willick J.A., Courteau S., Faber S.M., Burstein D., 
             Dekel A., 1995, ApJ, 446, 12
\reference{} Wolfe A., Turnshek, D.A., Smith, H.E., Cohen, R. D.,
             1986, ApJS, 61, 249
\reference{} Wolfe A., 1995,
             in QSO Absorption Lines, ed. G. Meylan, Springer: Berlin, p.13
\reference{} Zwaan M.A., van der Hulst J.M., de Blok W.J.G.,
             McGaugh S.S., 1995, MNRAS, 273, L35


\end{references}
\end{document}